\documentclass[11pt,a4paper]{article}
\usepackage{jcappub}
\usepackage{amssymb}
\usepackage{amsmath}
\newcommand{\be}{\begin{equation}}
\newcommand{\ee}{\end{equation}}
\newcommand{\bea}{\begin{eqnarray}}
\newcommand{\eea}{\end{eqnarray}}
\newcommand{\beaa}{\begin{eqnarray*}}
\newcommand{\eeaa}{\end{eqnarray*}}

\newcommand{\diag}{\mathop{\mathrm{diag}}\nolimits}

\title{Cosmological Solutions of a Nonlocal Model with a Perfect Fluid}
\author[a]{Emilio Elizalde,}
\affiliation[a]{Instituto de Ciencias del Espacio (ICE/CSIC) and
Institut d'Estudis Espacials de Catalunya (IEEC) Campus UAB,
Facultat de Ci\`encies, Torre C5-Parell-2a planta, 08193 Bellaterra
(Barcelona), Spain} \emailAdd{elizalde@ieec.uab.es}
\author[b]{Ekaterina O. Pozdeeva,}
\affiliation[b]{Skobeltsyn Institute of Nuclear Physics,  Lomonosov
Moscow State University, Leninskie Gory 1, 119991, Moscow, Russia}
 \emailAdd{pozdeeva@www-hep.sinp.msu.ru}
\author[b]{Sergey Yu. Vernov}
\emailAdd{svernov@theory.sinp.msu.ru}
\author[c]{and Ying-li Zhang}
\affiliation[c]{Yukawa Institute for Theoretical Physics, Kyoto
University, Kyoto 606-8502, Japan}
\emailAdd{yingli@yukawa.kyoto-u.ac.jp}

\abstract{ A nonlocal gravity model which does not assume the
existence of a new dimensional parameter in the action and includes
a function $f(\Box^{-1} R)$, with $\Box$ the d'Alembertian operator,
is studied. By specifying an exponential form for the function~$f$
and including a matter sector with a constant equation of state
parameter, all available power-law solutions in the Jordan frame are
obtained. New power-law solutions in the Einstein frame are also
probed. Furthermore, the relationship between power-law solutions in
both frames, established through conformal transformation, is
substantially clarified. The correspondence between power-law
solutions in these two frames is proven to be a very useful tool in
order to obtain new solutions in the Einstein frame.}


\arxivnumber{1302.4330}

\begin{document}
\maketitle
\section{Introduction}

The acceleration of the Universe expansion is presently supported by
a large number of independent sets of observational data, of very
different
kind~\cite{Perlmutter:1998np,Riess:1998cb,WMAP,Tegmark,Seljak:2004xh,Eisenstein:2005su,Jain:2003tba}.
Modern cosmological surveys allow astronomers to obtain increasingly
accurate joint constraints on the set of cosmological parameters
(see, e.g., \cite{Kilbinger:2008gk}). The usual assumption that
General Relativity (GR) is the correct theory of gravity at all
scales leads to the remarkable conclusion that about seventy per
cent of the energy density of the Universe at present must be
smoothly distributed under the form of a slowly varying cosmic fluid
with negative pressure, called dark energy. Ordinarily, in order to
specify a would be component of the cosmic fluid use is made of its
equation of state (EoS), namely a phenomenological relation between
the pressure, $p$, and the energy density, $\rho$, corresponding to
the considered component, e.g. $ p=w\rho$, where $w$ is the EoS
parameter. Contemporary experiments provide strong support to the
statement that the dark energy EoS parameter is presently very close
to $-1$. If this number were the exact value, this would lead us
back to GR with a cosmological constant (and nothing else), but a
small deviation from this value cannot be excluded by the most
accurate astronomical data available. Moreover, the sign (positive
or negative) or the tendency (e.g., the derivative) of this
deviation cannot be clearly determined at present. This makes room
for a number of theoretical models, derived from quite different
fundamental theories, which can accommodate such situation.

Actually, a few types of models exist which are able to reproduce
the observed late-time cosmic acceleration. The simplest, and most
popular of them is $\Lambda$CDM, which fits a wide range of
cosmological data~\cite{Seljak:2004xh}. In this model the dark
energy component is just the cosmological constant which is added to
the action corresponding to GR. Other models introduce a dynamical
dark energy characterized by a varying EoS parameter.
The standard way to obtain an evolving EoS parameter is the
addition of scalar fields to the cosmological model. Actually, the
evolution of the Universe is sufficiently well described by
cosmological models with scalar fields, in particular, by quintom
models, which involve two of them:
 a phantom scalar field and an ordinary scalar  (see e.g.~\cite{QuintomModels}). Quintom models
are being very actively studied at
present~\cite{Guo2004,AKV2,Andrianov,Lazkoz,Setare} (for reviews,
see also~\cite{Quinmodrev1}). We should note, however, that the
origin of this fluid, which produces anti-gravitational effects,
still remains a mystery. Other popular theories involve
modifications of Einsteinian gravity, as for instance $F(R)$
gravity, with $F(R)$ an (in principle) arbitrary function of the
scalar curvature $R$ (for recent reviews see,
e.g.,~\cite{NO-rev,Review-Nojiri-Odintsov,Felice_Tsujikawa,Fujii_Maeda,
Book-Capozziello-Faraoni,CL,BCNO}).

Higher-derivative corrections to the Einstein--Hilbert action are
being actively studied in the context of quantum gravity (as one of
the first papers on this subject we can mention~\cite{Stelle}; see
also~\cite{BOSh} and references therein). A nonlocal gravity theory
obtained by taking into account quantum effects was proposed
in~\cite{Deser:2007jk}. Also, string/M theory is usually considered
as a possible frame (expectedly, the ultimate one) for the
discussion of all fundamental interactions, including gravity,
consequently, the natural appearance of non-locality within string
field theory provides a very strong motivation for studying nonlocal
cosmological models. It should be emphasized in this context that
most of the nonlocal cosmological models available explicitly
include a function of the d'Alem\-bertian operator, $\Box$, and
either directly define a nonlocal modified
gravity~\cite{Non-local-gravity-Refs,Odintsov0708,Jhingan:2008ym,Koivisto:2008,Koivisto2,Woodard,
Non-local-FR,Tsamis:2010pt,ParkDonelson,Nojiri:2010pw,Bamba1104,ZS,EPV2011,Vernov1202,EPV2012}
or, alternatively, add a nonlocal scalar field, minimally coupled to
gravity~\cite{Non-local_scalar}.

In the present paper, we consider a nonlocal gravity model which
contains a function of the $\Box^{-1}$ operator but does not assume
the existence of a new dimensional parameter in the
action~\cite{Deser:2007jk}. For this kind of nonlocal models, an
explicit technique for choosing the distortion function,
$f(\Box^{-1})$, so as to fit an arbitrary expansion history, has
been derived in~\cite{Woodard}, and the specific nonlocal model
considered has a local scalar-tensor
formulation~\cite{Odintsov0708}. The perturbation analysis of this
model has been carried out and the Solar System test has been
performed in~\cite{Koivisto2}. De Sitter solutions and expanding
universe solutions with $a\sim t^n$ have been investigated
in~\cite{Odintsov0708,Bamba1104,ZS,EPV2011}. In~\cite{Koivisto:2008}
the ensuing cosmology describing the four basic epochs was studied
for nonlocal models involving, in particular, an exponential form
for the function $f(\eta)$. An explicit mechanism to screen the
cosmological constant in nonlocal gravity was discussed
in~\cite{Nojiri:2010pw,Bamba1104,ZS}. In the framework of the local
formulation, a reconstruction procedure was proposed
in~\cite{Koivisto:2008} and it has been developed
in~\cite{EPV2011,Vernov1202,EPV2012}.

 The example most usually
studied~\cite{Odintsov0708,Jhingan:2008ym,Koivisto:2008,Nojiri:2010pw,Bamba1104,ZS,EPV2011}
 of a model of this kind  is characterized by an exponential
function $f(\Box^{-1}R)=f_0 e^{\alpha(\Box^{-1}\!R)}$, where $f_0$
and $\alpha$ are real parameters, a case that will be explicitly
considered in this paper.

Conformal (Weyl) transformations are widely used in scalar-tensor
theories of gravity, the theory of a scalar field coupled
nonminimally to the Ricci curvature, $R$, and in $F(R)$ gravity
theories~\cite{BD:1961,Faraoni_OLD,Faraoni-Nadeau} (see also
\cite{Felice_Tsujikawa,Book-Capozziello-Faraoni,CL}). The
Hilbert--Einstein action and the modified gravity action can be
related by the conformal
transformation~\cite{Felice_Tsujikawa,Book-Capozziello-Faraoni,CL,Faraoni-Nadeau,CNOT,Capozziello,NS:2010a},
being the corresponding equations also connected by the same
transformation. The very important issue concerning which of the conformal
frames, Jordan or Einstein, is the true physical one, has been the
subject of longstanding debate
(see~\cite{Faraoni_OLD,Faraoni-Nadeau}, and references therein).  In
this respect, knowledge of the transitions between these frames is a
very useful tool for the construction of new exact solutions. We
prove this statement in Sect.~\ref{conform}.

The nonlocal model we will consider is usually studied in the Jordan
frame, but recently its behavior in the Einstein frame, for the
model without matter~\cite{Bamba1104,ZS}, has been explored too. In
this paper, we will focus our effort on the study of cosmological
solutions of this model, both in the Jordan and in the Einstein
frames, including the case with matter for the last one. We will
consider gravity models with a cosmological constant $\Lambda$ and
including a perfect fluid, and study in detail their cosmological
solutions with a power-law cosmic scalar factor: $a\propto t^n$. The
solutions thus obtained will be proven to generalize solutions
found in~\cite{Odintsov0708,Bamba1104,ZS}. In the Jordan frame we
find a class of power-law solutions and prove, moreover, that other
power-law solutions cannot be exact. We will analyze with care the
correspondence existing between the solutions obtained in the
different frames and will demonstrate explicitly how the explicit knowledge
of power-law solutions in the Jordan frame can be used in order to
get power-law solutions in the Einstein frame.

The paper is organized as follows. In Sect.~\ref{origin}, we start
from the action for a general class of nonlocal gravity, without
specifying the form of the function $f(\psi)$, and derive the
equations of motion for a spatially flat cosmology in the Jordan
frame. The theory is specialized in Sect.~\ref{jordan} to a model
characterized by the function $f(\psi)=f_0e^{\alpha\psi}$, and the
corresponding power-law solutions are obtained. In Sect.~\ref{Nonlocal},
we consider the initial nonlocal model with a perfect fluid. Using the power-law solutions
of Sect.~\ref{jordan}, we then get the class of power-law solutions for this nonlocal model.
In Sect.~\ref{JE} we investigate a conformal transformation from the
original (Jordan) to the Einstein frame, and derive the
corresponding equations of motion (EOM). Vacuum power-law solutions in the
Einstein frame are derived in Sect.~\ref{Einvacuum}. In
Sect.~\ref{einsolutions}, we include the matter sector and obtain
the corresponding power-law solutions by directly solving the EOM.
In Sect.~\ref{conform} we use the correspondence between power-law
solutions in the Jordan and in the Einstein frames in order to get
brand new solutions in the Einstein frame. Finally, Sect.~\ref{conclude} is
devoted to conclusions.

\section{Nonlocal gravitational action and the equations of motion}
\label{origin} We start by considering a class of nonlocal
gravities, with action given by
\begin{equation}
\label{action} S=\int d^4 x \sqrt{-g}\left\{
\frac{1}{2\kappa^2}\left[ R\left(1 + f(\Box^{-1}R)\right) -2\Lambda
\right] + \mathcal{L}_\mathrm{m} \right\}\, ,
\end{equation}
where $\kappa^2=8\pi G=8\pi/{M_{\mathrm{Pl}}}^2$, the Planck mass
being $M_{\mathrm{Pl}} = G^{-1/2} = 1.2 \times 10^{19}$ GeV, while
$f$ is a differentiable  function, which characterizes the nature of
nonlocality, $\Box^{-1}$ being the inverse of the d'Alembertian
operator, $\Lambda$ the  cosmological constant, and
$\mathcal{L}_\mathrm{m}$ the matter Lagrangian. For definiteness, we
assume that matter is a perfect fluid.  We use the signature
$(-,+,+,+)$, $g$ being the determinant of the metric tensor,
$g_{\mu\nu}$. Recall the covariant d'Alembertian for a scalar field,
which reads
\begin{equation*}
\Box\equiv \nabla^\mu\nabla_\mu= \frac{1}{\sqrt{-g}} \partial_\mu
\left( \sqrt{-g} \, g^{\mu
  \nu}\partial_\nu \right),
\end{equation*}
where $\nabla^\mu$ is the covariant derivative.

For practical uses, introducing two scalar fields
$\psi=\Box^{-1}R$ and a Lagrange multiplier $\xi$
(see~\cite{Odintsov0708}), action~(\ref{action}) can be recast as a
local action, namely
\begin{equation}
\label{action2} S_l= \int d^4x \sqrt{-g}\left\{
\frac{1}{2\kappa^2}\left[ R\left(1 + f(\psi)\right) +\xi
\left(R-\Box \psi\right) - 2 \Lambda \right]+ \mathcal{L}_\mathrm{m}
\right\} \, .
\end{equation}
Therefore, the original action can actually be regarded as a local
one (\ref{action2}) in the Jordan frame. By varying this action with
respect to $\xi$ and $\psi$, one respectively gets the field
equations \bea \label{equpsi} \Box\psi&=&R\,,
\\
\Box\xi&=& f_{,\psi}(\psi) R\, , \label{equxi} \eea where
$f_{,\psi}(\psi)\equiv{\rm{d}}f/{\rm{d}\psi}$. The corresponding
Einstein equations are obtained by variation of the
action~(\ref{action2}) with respect to the metric tensor
$g_{\mu\nu}$, as follows
\begin{equation}
\label{nl4}
\begin{split}
&\frac{1}{2}g_{\mu\nu} \left[R\Psi
 + \partial_\rho \xi \partial^\rho \psi - 2 (\Lambda+\Box\Psi) \right]
 - R_{\mu\nu}\Psi-\frac{1}{2}\left(\partial_\mu \xi \partial_\nu \psi
+ \partial_\mu \psi \partial_\nu \xi \right)
  + \nabla_\mu \partial_\nu\Psi=- \kappa^2T_{(\mathrm{m})\, \mu\nu}\, ,
\end{split}
\end{equation}
where $\Psi\equiv1+f(\psi) +\xi$, and $T_{(\mathrm{m})\,\mu\nu}$ is
the energy--momentum tensor of the matter sector, defined as
\begin{equation}\label{Tmatter}
T_{(\mathrm{m})\, \mu\nu} \equiv -\frac{2}{\sqrt{-g}}
 \frac{\delta \left(\sqrt{-g} \mathcal{L}_\mathrm{m}\right)}{\delta g^{\mu \nu}}.
\end{equation}
We note that the system of equations here considered does not
include the function $\psi$ itself, but instead $f(\psi)$ and
$f_{,\psi}(\psi)$, together with time derivatives of $\psi$. Also,
$f(\psi)$ can only be determined up to a constant, since one may
indeed add a constant to $f(\psi)$ and subtract the same constant
from $\xi$ without changing the original equations at all.

In this paper, we assume a spatially flat
Friedmann--Lema\^{i}tre--Ro\-bertson--Walker (FLRW) universe, with
the space-time interval
\begin{eqnarray}\label{FLRW}
ds^2={}-dt^2+a^2(t)\delta_{ij}dx^idx^j\,,
\end{eqnarray}
and consider the case where the scalar fields $\psi(t)$ and $\xi(t)$
are only functions of the cosmological time. Thus, the system of
Eqs.~(\ref{equpsi})--(\ref{nl4}) reduces to
 \bea \label{einstein1} 3
H^2\Psi&=& \!{}- \frac{1}{2}\dot\xi \dot\psi
 - 3H\dot\Psi  + \Lambda
+ \kappa^2 \rho_{\mathrm{m}}\, ,\\
\label{einstein2} \left(2\dot H + 3H^2\right)\Psi &=&
\frac{1}{2}\dot\xi \dot\psi -
\ddot\Psi- 2H \dot\Psi + \Lambda - \kappa^2 P_{\mathrm{m}}\,,\\
\label{psieq}\ddot \psi &=&  \!{}- 3H \dot \psi - 6 \left( \dot H +
2
H^2\right) \, , \\
\label{xieq} \ddot \xi  &=& \!{}- 3H \dot \xi - 6\left( \dot H + 2
H^2\right)f_{,\psi}(\psi) \,, \eea where a dot means differentiation
with respect to time, $t$, in the Jordan frame: $\dot
{A}(t)\equiv{\rm{d}}A(t)/{\rm{d}}t$, and $H=\dot a/a$ is the Hubble
parameter. For a perfect matter fluid, we have $T_{(\mathrm{m})\, 0
0} = \rho_{\mathrm{m}}$ and $T_{(\mathrm{m})\, i j} = P_{\mathrm{m}}
g_{i j}$. The continuity equation is
\begin{equation}
\label{equ_rho} \dot\rho_{\mathrm{m}}={}-
3H(P_{\mathrm{m}}+\rho_{\mathrm{m}}).
\end{equation}
It is useful to add up (\ref{einstein1}) and (\ref{einstein2}), and
get
\begin{equation}
\label{equPsi} \ddot\Psi+5H\dot\Psi+\left(2\dot H + 6H^2\right)\Psi-
2\Lambda +\kappa^2 (P_{\mathrm{m}}-\rho_{\mathrm{m}})=0.
\end{equation}
Note that Eq.~(\ref{equPsi}) is a second-order linear differential
equation for~$\Psi$.

\section{Power-law solutions of the model with $f(\psi)$ an exponential function}
\label{jordan}
\subsection{The model with $f(\psi)$ being an exponential function}
Following~\cite{Bamba1104,ZS,EPV2011}, we  consider the case where
$f(\psi)$ is an exponential function, namely
\begin{equation}
\label{f} f(\psi)=f_0 e^{\alpha\psi}\, ,
\end{equation}
with $f_0$ and $\alpha$ nonzero real parameters. The motivation for
considering  an exponential function $f(\psi)$ is not only because
it is the simplest model with power-law and de Sitter
solutions\footnote{In models with such solutions,  the function
$f(\psi)$ is either an exponential function or a sum of exponential
functions~\cite{EPV2012}.}, but also, because it is the better
studied case among all possible functions
$f(\psi)$~\cite{Odintsov0708,Jhingan:2008ym,Koivisto:2008,Nojiri:2010pw,Bamba1104,ZS,EPV2011}
(de Sitter solutions for this model were discussed
in~\cite{Odintsov0708,Bamba1104,EPV2011}, and expanding universe
solutions with the Hubble parameter $H=n/t$, where $n$ is a nonzero
constant, in~\cite{Odintsov0708,ZS}). In the present
paper we will investigate this last type of solutions in detail.

We consider matter with the EoS
parameter $w_{\mathrm{m}}\equiv P_{\mathrm{m}}/\rho_{\mathrm{m}}$
being a constant but {\it not} equal to $-1$. For power-law
solutions $H=n/t$, Eq.~(\ref{equ_rho}) has the following general
solution:
\begin{equation}
\label{rho_sol}
\rho_{\mathrm{m}}(t)=\rho_0t^{-3n(w_{\mathrm{m}}+1)},
\end{equation}
where $\rho_0$ is an arbitrary constant.

\subsection{Solutions with $H=n/t$}

The goal of this section is to find the whole set of power-law
solutions for the model, characterized by the function $f$ given in
(\ref{f}). In this subsection, we present some power-law solutions
and, in the next one, we will show that no other power-law solutions
exist.

Inserting $H=n/t$ into  Eq.~(\ref{psieq}), the following solution
$\psi(t)$ is obtained,
\begin{equation}
\label{psisoln}
\psi(t)=\psi_1t^{1-3n}-\frac{6n(2n-1)}{3n-1}\ln\left(\frac{t}{t_0}\right)\,,
\end{equation}
where $\psi_1$  and $t_0$ are integration constants. We consider real solutions at $t>0$,
hence, $t_0>0$. Note that this
solution is valid provided $n\neq1/3$ and $n\neq1/2$. Consequently,
in this subsection the cases $n=1/2$ and $n=1/3$ will be excluded
from our analysis. We also, specify $\psi_1=0$, so that the function
$f(\psi)$ takes the following form
\begin{equation}\label{fm}
f(\psi(t))=f_0\left(\frac{t}{t_0}\right)^m\,, \qquad
m\equiv{}-6\alpha \frac{ n(2n-1)}{3n-1}\,.
\end{equation}
We will show in the next subsection that there is no solution for
$\psi_1\neq 0$. The cases $n=1/2$ and $n=1/3$ will be considered in
Sect.~\ref{SpecialCases}.

Inserting formulae~(\ref{psisoln}) and (\ref{fm}) into (\ref{xieq})
 one obtains the following expression for~$\xi(t)$
\begin{equation}
\label{xisoln} \xi(t)=\left\{ \begin{aligned}
         &\xi_0+\xi_1\left(\frac{t}{t_0}\right)^{1-3n} +\frac{(3n-1)f_0}{3n+m-1}\left(\frac{t}{t_0}\right)^{m}\,,\quad \mbox{for } m\neq1-3n, \\
         &\xi_2-mf_0\left(\frac{t}{t_0}\right)^m\ln\left(\frac{t}{t_1}\right)\,,\quad \mbox{for }
         m=1-3n,\\
                          \end{aligned} \right.
\end{equation}
where $\xi_0$, $\xi_1$, $\xi_2$, and $t_1$ are integration
constants.

Furthermore, substituting the solutions described by formulae
(\ref{rho_sol}), (\ref{psisoln}), and  (\ref{xisoln}), into
Eqs.~(\ref{einstein1}) and (\ref{einstein2}), constraints on these
integration constants can be obtained
\begin{itemize}
\item For $ m\neq 1-3n$, which equivalently implies the following constraint on the power index~$n$
\begin{equation}\label{nneq}
n\neq
\frac{3(\alpha-1)\pm\sqrt{3\alpha(3\alpha-2)}}{3(4\alpha-3)}\,,
\end{equation}
in this case, we get constraints on the integration constants for
$\Lambda=0$ and $\Lambda\neq0$ separately:
\begin{itemize}
\item
For $\Lambda=0$, by inserting the solutions (\ref{rho_sol}),
(\ref{psisoln}), and (\ref{xisoln}) into the system
(\ref{einstein1})--(\ref{xieq}), the corresponding integration
constants are fixed by
\begin{equation}
\label{constantvalues} \left\{ \begin{aligned}
         \xi_0 &= -1\,,\\
         \rho_0 &=
\frac{6(3n-1+3\alpha-6n\alpha)f_0n^2}{(3n-1)\kappa^2}t_0^{6n(2n-1)
\alpha/(3n-1)}\, ,
                          \end{aligned}
\right.
\end{equation}
with $t_0$ and $\xi_1$ to be determined by initial conditions, while
the power index $n$ is constrained by
\begin{equation}
\label{equ_wn}
w_{\mathrm{m}}+1-\frac{2}{3n}-\frac{2\alpha (2n-1)}{3n-1}=0,
\end{equation}
from which $n$ is expressed in terms of the parameters $w_\mathrm{m}$
and $\alpha$:
\begin{equation}\label{nbyw}
n=n_\pm=\frac{3w_\mathrm{m}-6\alpha+9\pm\sqrt{\left(3w_\mathrm{m}-6\alpha+1\right)^2
+8\left(1-3w_\mathrm{m}\right)}}{6(3w_\mathrm{m}-4\alpha+3)}\,.
\end{equation}
This furthermore yields corresponding constraints on the parameters
$\alpha$ and $w_\mathrm{m}$ for a real number $n_\pm$:
\begin{equation}\label{alpharange}
\left(3w_\mathrm{m}-6\alpha+1\right)^2+8\left(1-3w_\mathrm{m}\right)\geqslant0\,.
\end{equation}

Interestingly enough, from Eq.~(\ref{nbyw}) one finds that, for
$\alpha\gg1$, one of the power indices behaves as
$n_-\longrightarrow1/2$, which implies that the Universe
asymptotically evolves to a radiation-dominated phase, and this
regardless of the details of the EoS for the matter sector. Thus, we
manage to obtain an expanding universe without introducing a
cosmological constant $\Lambda$. This is because the non-local term
$\Box^{-1}R$ plays partially the role of a dark energy, though it is
a decelerating expansion.

Furthermore, we pay attention to two special values of the EoS
parameter:

(i) When $w_\mathrm{m}=-1$  matter is nothing but just an effective
cosmological constant. Therefore, this case corresponds to
$\Lambda\neq 0$.

(ii) If the only matter is radiation, namely, $w_{\mathrm{m}}=1/3$,
Eq.~(\ref{nbyw}) leads to $n_-=1/2$, which should be excluded.
Therefore, the power index is $n=n_+$ for the radiation component.

\item
For $\Lambda\neq 0$, the corresponding integration constants and
parameters are constrained:
\begin{equation}\label{nonvanishlamb}
\left\{ \begin{aligned}
         \quad m&=2\,,\\
t_0^2&=\frac{6n(n+1)f_0}{\Lambda}, \,\\
         \rho_0 &= \frac{3(1+\xi_0)n^2}{\kappa^2}\,,
\end{aligned}
\right.
\end{equation}
the integration constants $\xi_0$ and $\xi_1$ in the solution (\ref{xisoln}) are to be fixed
by  the initial conditions, while the power-index $n$ is determined
by $w_\mathrm{m}$:
\begin{equation}
\label{nwm} n=\frac{2}{3(1+w_\mathrm{m})}\,.
\end{equation}
Here we note that, by recalling the definition of $m$ in
Eq.~(\ref{fm}) and using Eq.~(\ref{nwm}), one finds that the
parameter $\alpha$ is constrained by $m=2$:
\begin{equation}
\label{equn}
\alpha=\frac{3(1-w_\mathrm{m}^2)}{2(3w_\mathrm{m}-1)}\,.
\end{equation}
Thus, unlike the situation for vacuum solutions with a nonzero
cosmological constant in~\cite{ZS}, here we find that with a matter
sector the parameter of the model $\alpha$ is fixed by the EoS of
matter in Eq.~(\ref{equn}). In this sense, the model is spoiled
since, in this case, it cannot yield a smooth evolution of the
Universe for different stages with a given parameter $\alpha$.

 From (\ref{equn}) it follows that there is no
power-law solution for non-vanishing cosmological constant with
matter whose EoS is $w_\mathrm{m}=1/3$ or $w_\mathrm{m}=1$, if $m\neq 1-3n$. The case
$w_\mathrm{m}=-1$ corresponds to the cosmological constant as the matter
part. Hence, to obtain the solutions in this case we should put
$\rho(t)=0$, which corresponds to $\xi_0=-1$ from
Eq.~(\ref{nonvanishlamb}). In this case, from the constraint $m=2$,
we also obtain two branches for the power-index:
\begin{equation}
\label{n_alpha} n=n_\pm
=\frac{3(\alpha-1)\pm\sqrt{9\alpha^2+6\alpha+9}}{12\alpha}\,,
\end{equation}
which coincide with Eq.~(25) of Ref.~\cite{ZS}, as expected. So the
solutions obtained here contain the vacuum case where the universe
asymptotically behaves like a radiation-dominated one for large
$\alpha$.

\end{itemize}
\item For $m=1-3n$, the power-index $n$ is determined\footnote{The equation is the following:
\begin{equation*}
 \alpha-\frac{(3n-1)^2}{6n(2n-1)}=0.
\end{equation*}} by the parameter $\alpha$:
\begin{equation}\label{nalphaequ}
n=n_\pm=\frac{3(\alpha-1)\pm\sqrt{3\alpha(3\alpha-2)}}{3(4\alpha-3)}\,.
\end{equation}

For $\Lambda=0$, one finds power-law solutions with the following
constraints on the integration constants
\begin{equation}
\left\{ \begin{aligned}
         \xi_2& = {}-1\,,\\
         \rho_0 &={}-\frac{3n(n-1)f_0}{\kappa^2}t_0^{3n-1}\,,
\end{aligned} \right.
\label{coeff3}
\end{equation}
while, again, $n$ is determined by $w_\mathrm{m}$
\begin{equation}\label{nm2}
n=\frac{1}{3w_\mathrm{m}}\,;
\end{equation}
thus, similarly as in the former case for $m\neq1-3n$ with nonzero
cosmological constant $\Lambda$, here the parameter $\alpha$ is also
fixed by $w_\mathrm{m}$
\begin{equation}\label{alphm2}
\alpha=\frac{3(1-w_\mathrm{m})^2}{2(2-3w_\mathrm{m})}\,,
\end{equation}
which again implies that in this case the model cannot yield the
different stages of the Universe evolution with a fixed $\alpha$.

For some special values of the parameter $\alpha$
additional solutions exist, namely:
\begin{itemize}
\item
For $\alpha=2/3$ and $\Lambda=0$, there exists a solution with
integral constants $\xi_2$, $t_0$, and $t_1$ to be specified by the
initial conditions, while $n$, $w_\mathrm{m}$ and $\rho_0$ are fixed by
\begin{equation}
 \qquad n=1,\qquad w_{\mathrm{m}}={}-\frac13\,,\qquad
\rho_0=\frac{3(\xi_2+1)}{\kappa^2}.
\end{equation}
\item
When $\alpha=6/5$ and $\Lambda\neq 0$ we obtain a solution with
$t_1$ undetermined, while
\begin{equation}\label{al65}
n={}-\frac{1}{3},\qquad t_0^2= {}-\frac{4f_0}{3\Lambda}\,, \qquad
\rho_0=0\,,\qquad  \xi_2={}-1.
\end{equation}
\end{itemize}
Note that for $\alpha=6/5$ we also have a solution (\ref{coeff3})
with $\Lambda=0$ and $n = 5/9$.
\end{itemize}

Thus, we have obtained corresponding solutions for both nonzero and
zero values of $\Lambda$. They generalize the ones previously found
in the absence of matter~\cite{ZS}. All solutions for the case
$m=1-3n$ are new.

\subsection{Proof of the absence of power-law solutions in the case $\psi_1\neq 0$}

Recall at this point that our main goal is to find all solutions that
correspond to $H=n/t$. In this subsection we consider the case
$\psi_1\neq 0$ with the hope to obtain new solutions or else rigorously
prove that such solutions do not exist.

In the case $\psi_1\neq 0$, Eq.~(\ref{xieq}) has no solution in
terms of elementary functions, thus it is more convenient to solve
Eq.~(\ref{equPsi}), get $\Psi(t)$, and substitute
$\xi(t)=\Psi(t)-f(\psi(t))$ into Eq.~(\ref{xieq}), to then check if
there exist values of the parameters for which the obtained $\xi(t)$
satisfies Eq.~(\ref{xieq}), or not. The type of solutions of
Eq.~(\ref{equPsi}) depends on the value of $n$, thus we consider
different cases.

For $n\neq -1$ and $n \neq -1/3$, Eq.~(\ref{equPsi}) has the
following general solution
\begin{equation}
\label{solPsi1}
\Psi(t)=C_1t^{-2n}+C_2t^{1-3n}+\frac{\Lambda}{(n+1)(3n+1)}
t^2-\frac{\rho_0
\kappa^2(w_{\mathrm{m}}-1)t^{2-3(1+w_{\mathrm{m}})n}}{(3nw_{\mathrm{m}}-1)(n+3nw_{\mathrm{m}}-2)},
\end{equation}
where $C_1$ and $C_2$ are integral constants and $w_{\mathrm{m}}$ is
chosen so that $(3nw_{\mathrm{m}}-1)(n+3nw_{\mathrm{m}}-2)$ be not
equal to zero.

In the case $n\neq - 1$ and $n\neq - 1/3$, the solutions of
Eqs.~(\ref{psieq}) and (\ref{equPsi}) are given by
formulae~(\ref{solPsi1}) and (\ref{psisoln}), respectively.
Substituting the $\xi(t)=\Psi(t)-f(\psi(t))$ obtained into
(\ref{xieq}), we realize that this equation is not satisfied. In
particular, the non-matching expression is proportional to
\begin{equation*}
t^{\gamma}\exp\left(-\frac{\psi_1}{\beta} t^{1-3n}\right),
\end{equation*}
where $\beta$ and $\gamma$ are combinations of constants. The
exponential term disappears only for $\psi_1=0$. Therefore, there is
no solution for $\psi_1\neq 0$ and $n\neq - 1$, $n\neq - 1/3$.
Similar calculations show that, in the cases $n=-1$ and $n= - 1/3$,
solutions are absent, as well.

We thus have found all solutions which correspond to $H=n/t$. Note
that, in contradistinction to the papers~\cite{Odintsov0708,ZS}, we
here include an ideal perfect fluid in action (\ref{action2})
and do not impose any restrictions whatsoever on the parameters and integration
constants.

\subsection{Special values of the power index $n$}
\label{SpecialCases}

Let us consider the case $n=1/2$, which corresponds to $R=0$.
Solving Eqs.~(\ref{psieq}) and (\ref{xieq}), we get
\begin{equation}\label{triv1}
\psi(t)=\psi_3t^{-1/2}+\psi_4, \qquad \xi(t)=\xi_3t^{-1/2}+\xi_4,
\end{equation}
where $\psi_3$, $\psi_4$, $\xi_3$, and $\xi_4$ are integral
constants. Straightforward substitution of these functions and
$H=1/(2t)$ into Eqs.~(\ref{einstein1}) and (\ref{einstein2}) yields a
solution for $\Lambda=0$ with the following conditions on the
constants:
\begin{equation}\label{triv2}
\psi_3=0,\qquad \xi_4 =
-1-f_0e^{\alpha\psi_4}+\frac{4}{3}\kappa^2\rho_0, \qquad
w_{\mathrm{m}}=\frac{1}{3},
\end{equation}
while $\rho_0$, $\psi_4$ and $\xi_3$ are to be determined by the
initial conditions.

When $n=1/3$, Eq.~(\ref{psieq}) has the solution:
\begin{equation} \psi(t) =
\frac{1}{3}\ln\left(\frac{t}{t_2}\right)^2+\psi_5\ln\left(\frac{t}{t_2}\right),
\end{equation}
where $\psi_5$ and $t_2$ are integration constants. The function
$\xi(t)$, as a solution of  Eq.~(\ref{xieq}), can be given in terms of
quadratures only. At the same time, solving Eq.~(\ref{equPsi}), we
get $\Psi(t)$ in terms of elementary functions. Thus, if a solution
exists, then the corresponding $\xi(t)$ should be an elementary
function as well. We therefore arrive to a contradiction, what
proves the absence of power-law solutions with $n=1/3$.

\subsection{Brief summary of the solutions in the Jordan frame}
\label{jordantable}
To make it easier for readers to look at the
whole set of solutions, we list all those we have found in
this section in Tables~\ref{t1f1} and \ref{t1f2}. We note that, in
the case $n\neq1/2,~1/3$, the solution for $\psi(t)$ is uniquely
given by Eq.~(\ref{psisoln}) with $\psi_1=0$, i.e.
\begin{equation}
\label{psisoln1}
\psi(t)=-\frac{6n(2n-1)}{3n-1}\ln\left(\frac{t}{t_0}\right)\,,
\end{equation}
and this expression is not repeated in the table. It should be
noted that for $n=1/2$ the solution is given by Eqs.~(\ref{triv1})
and (\ref{triv2}), while no power-law solution exists for $n=1/3$.

\begin{table}[tbp]
\centering
\begin{tabular}{l@{}ll@{}ll@{}l}
\hline
\multicolumn{2}{c}{$m\neq1-3n$}&\multicolumn{2}{c}{solutions}&\multicolumn{2}{c}{constraints}
\\ \hline
&&&&\\
$\Lambda=0$& &$\begin{cases}&{\displaystyle
\xi(t)=-1+\xi_1\left(\frac{t}{t_0}\right)^{1-3n} +\frac{(3n-1)
f_0}{3n+m-1}\left(\frac{t}{t_0}\right)^{m}}
\,,\\
\\
&{\displaystyle\rho_{\mathrm{m}}(t)=\frac{6f_0n^2}{\kappa^2t_0^2}\left[1+\frac{3\alpha(1-2n)}{3n-1}\right]
\left(\frac{t}{t_0}\right)^{-3n(w_{\mathrm{m}}+1)}}\,,\end{cases}$& &$\begin{cases} \ \mbox{\rm Eq.}\, (\ref{equ_wn})\\ \ \mbox{\rm Eq.}\, (\ref{nbyw})\\ \ \mbox{\rm Eq.}\, (\ref{alpharange})\end{cases}$&  \\
&&&&\\
$\Lambda\neq0$&
&$\begin{cases}&{\displaystyle\xi(t)=\xi_0+\xi_1\left(\frac{t}{t_0}\right)^{1-3n}
+\frac{(3n-1)
f_0}{3n+1}\left(\frac{t}{t_0}\right)^{2}}\,,\\
\\
&{\displaystyle\rho_{\mathrm{m}}(t)=\frac{3n^2(1+\xi_0)}{\kappa^2}t^{-3n(w_{\mathrm{m}}+1)}}\,,
\end{cases}$&
&$\begin{cases} \ \mbox{\rm Eq.}\, (\ref{nonvanishlamb})\\ \ \mbox{\rm Eq.}\, (\ref{nwm})\\ \ \mbox{\rm Eq.}\, (\ref{equn})\end{cases}$\\
&&&&\\ \hline
\end{tabular}
\caption{Solutions in the Jordan frame for $m\neq1-3n$}\label{t1f1}
\end{table}

\begin{table}[tbp]
\centering
\begin{tabular}{l@{}ll@{}ll@{}l}
\hline
\multicolumn{2}{c}{$m=1-3n$}&\multicolumn{2}{c}{solutions}&\multicolumn{2}{c}{constraints}
\\ \hline
&&&&\\
$\Lambda=0$& &$\begin{cases}&{\displaystyle\xi(t)=-1+f_0(3n-1)\left(\frac{t}{t_0}\right)^{1-3n}\ln\left(\frac{t}{t_1}\right)}\,,\\
\\
&{\displaystyle\rho_{\mathrm{m}}(t)=\frac{3f_0n(1-n)}{\kappa^2t_0^2}{\left(\frac{t}{t_0}\right)^{-3n-1}}}\,,\end{cases}$&
&$\begin{cases} \ \mbox{\rm Eq.}\, (\ref{nalphaequ})\\ \ \mbox{\rm
Eq.}\, (\ref{nm2})\\ \ \mbox{\rm Eq.}\, (\ref{alphm2})\end{cases}$&
\\
&&&&\\
$\Lambda\neq0$&
&$\begin{cases}&{\displaystyle\xi(t)=-1+\frac{3\Lambda
t^2}{2}\ln\left(\frac{t}{t_1}\right)}\,,\\&\rho_m(t)=0\,,\end{cases}$&
&$\begin{cases}{\displaystyle \, \alpha=6/5}\\ \ \mbox{\rm Eq.}\, (\ref{al65})\end{cases}$&  \\
\hline
\end{tabular}
\caption{Solutions in the Jordan frame for $m=1-3n$}\label{t1f2}
\end{table}

\subsection{Local constraints}

Modified gravity theories are quite strictly constrained by local
observations~\cite{Felice_Tsujikawa,Will:2005va}.  Precise consideration~\cite{Koivisto2} of the Newtonian limit of the theory, described by action~(\ref{action2}),
gives the following restrictions on the post-Newtonian parameter $\gamma$:
\begin{equation}\label{PNconstr}
\left|\gamma-1\right|=\left|\frac{4f_{,\psi}}{1+f+\xi-8f_{,\psi}}\right|<2.3\times10^{-5}\,.
\end{equation}
In order to check whether the power-law solutions found in the previous
sections can satisfy this constraint, we choose the solution where
$m\neq1-3n$ and $\Lambda=0$, since this one may be most
relevant to the deceleration expansion phase when matter fields
dominate the expansion of the Universe. Hence, we take the
corresponding solution
\begin{equation} \label{eq:power}
         \psi(t)={}-\frac{6n(2n-1)}{3n-1}\ln\left(\frac{t}{t_0}\right),\qquad
         \xi(t)=\xi_1\left(\frac{t}{t_0}\right)^{1-3n}-1 +\frac{(3n-1)
f_0}{3n+m-1}\left(\frac{t}{t_0}\right)^{m}
                         \end{equation}
and obtain
\begin{eqnarray}\label{PNconstrpower}
\gamma-1=\frac{4f_{,\psi}}{1+f+\xi-8f_{,\psi}}=\frac{4f_0\alpha}{\xi_1
\left(\frac{t}{t_0}\right)^{1-3n-m}-f_0\left(8\alpha-1-\frac{3n-1}{3n+m-1}\right)}\,.
\end{eqnarray}
Now, we discuss whether the constraint Eq.~(\ref{PNconstr}) can be
satisfied or not, in each of the two different cases:

\begin{itemize}
\item $\xi_1\neq0$. \ The restrictions on the parameter $\gamma$ have been obtained by the consideration of the effects within the Solar System~\cite{Will:2005va}, so we can assume that $t$ is not small and that $t_0\ll t$. Provided $1-3n-m>0$, the constraint~(\ref{PNconstr}) can be easily
fulfilled. For example, supposing that the power index $n>1/2$, it translates into the following constraint on the parameter $\alpha$:
\begin{eqnarray}
1-3n-m>0\quad\Longrightarrow\quad
\alpha>\frac{(3n-1)^2}{6n(2n-1)}\,.
\end{eqnarray}

\item $\xi_1=0$. \  In this case Eq.~(\ref{PNconstrpower}) reduces to:
\begin{equation}\label{gam1}
\gamma=1-\frac{4\alpha}{8\alpha-1-\left(1+\frac{m}{3n-1}\right)^{-1}}\,,
\end{equation}
where
\begin{equation}
\frac{m}{3n-1}=-\frac{6\alpha n(2n-1)}{(3n-1)^2}\,.
\end{equation}
Without loss of generality, we assume $n\sim\mathcal{O}(1)$. It is convenient to divide the discussion into three cases:

1) $\quad$ $|\alpha|\gg1$. \  In this case, $|m/(3n-1)|\gg1$, hence,
\begin{equation}
|\gamma-1|\simeq\frac{1}{2}\,,
\end{equation}
and the local constraint (\ref{PNconstr}) cannot be satisfied.

2) $\quad$ $|\alpha|\ll1$. \  In this case, $|m/(3n-1)|\ll1$, hence,
\begin{equation}
|\gamma-1|\simeq|\alpha|<10^{-5}\,,
\end{equation}
which implies that we need to tune $\alpha$ to a very small value.

3) $\quad$ $|\alpha|\sim\mathcal{O}(1)$. \  In this case, we should
recall that the power-index $n$ is related to $\alpha$ and the EoS
parameter, $w_\mathrm{m}$, by Eq.~(\ref{nbyw}).

An especially interesting case is when the matter sector is composed
of a non-relativistic matter fluid, i.e. $w_\mathrm{m}=0$. In this
case, inserting $n_-$ into Eq.~(\ref{gam1}), one finds that we need
to specify $\alpha$ to be of order $10^{-5}$ for the local
constraint (\ref{PNconstr}). If we take the $n_+$ branch, besides
$|\alpha|\lesssim10^{-5}$, there is another point
$\alpha\approx0.75$, but the allowed range around this value is
about $10^{-5}$ for local constraints. Moreover, similar conclusions
hold for radiation components, for which the EoS parameter
$w_\mathrm{m}=1/3$.
\end{itemize}

Thus, depending on whether the integration constant $\xi_1$ is
non-vanishing or not, we draw different conclusions concerning the
constraint of the Post-Newtonian parameter $\gamma$: when $\xi_1\neq0$,
the constraint can be easily satisfied for a wide range of choices of
the parameter $\alpha$ in this model, but when $\xi_1=0$ one needs to tune
the parameter $\alpha$ to at least  $10^{-5}$ order, to satisfy the
local constraint.  Note that in previous papers~\cite{Odintsov0708,ZS} the authors just set
$\xi_1=0$ for simplicity. The analysis of the local constraint shows that
solutions with nonzero $\xi_1$ allow to change the restrictions on the
parameter $\alpha$, which are indeed necessary in order to make the model compatible
with astronomical observations.

\section{Power-law solutions for the original nonlocal model}
\label{Nonlocal}

In this section, we discuss the power-law solutions for the nonlocal model
in the original form (\ref{action}).
When we vary the
nonlocal action (\ref{action}) with respect to the metric
$g_{\mu\nu}$, under the spatially flat FLRW metric (\ref{FLRW}), the
independent components of field equations can be expressed
as follows~\cite{Tsamis:2010pt,ParkDonelson}:
\begin{equation}\label{nonlocal_equations}
\begin{split}
3H^2 + \Delta G_{00} &= \kappa^2 \rho_{\mathrm{m}}+\Lambda, \\
-2\dot{H} -3H^2 + \frac{1}{3a^2}\delta^{ij}\Delta
G_{ij} &= \kappa^2 P_{\mathrm{m}}-\Lambda,
\end{split}
\end{equation}
where $\Delta G_{00}$ and $\Delta G_{ij}$ denote the modifications
coming from the nonlocal terms, namely
\begin{eqnarray*}
\Delta G_{00} &=& \left[ 3H^2 + 3H\partial_t\right]
\left\{ f \Bigr( \Box^{-1}R\Bigl) + \Box^{-1}\Bigl[ {R}
\frac{df}{d(\Box^{-1}R)}\Bigr]\right\}
\\
&+&\frac{1}{2}
\partial_{t}\Bigr( \Box^{-1}R\Bigl)
\partial_{t}\bigg(\Box^{-1}\Bigl[ {R} \frac{df}{d(\Box^{-1}R)}\Bigr]\bigg)\;, \\
\Delta G_{ij} &=& a^2\delta_{ij}\left[\frac{1}{2}
\partial_{t}\Bigr( \Box^{-1}R\Bigl)
\partial_{t}\bigg(\Box^{-1}\Bigl[ {R} \frac{df}{d(\Box^{-1}R)}\Bigr]\bigg)\;\right.\\
&\!\!\!-\!\!\!&\left.\left[ 2\dot{H}+ 3H^2+2H\partial_t
+ \partial^2_t\right] \left\{ f \Bigr( \Box^{-1}R\Bigl) + \Box^{-1}\Bigl[ {R} \frac{df}{d(\Box^{-1}R)}\Bigr]\right\}\right] .
\end{eqnarray*}
Hence, the identification of the scalar fields $\psi$ and $\xi$ with corresponding terms in the
original action yields~\cite{Woodard}
\begin{eqnarray}
\psi(t)&=&\Box^{-1}R\,,\label{psiidenti}\\
\xi(t)&=&\Box^{-1}\left[R\frac{df}{d(\Box^{-1}R)}\right]\,.\label{xiidenti}
\end{eqnarray}
Substituting these expressions into the system (\ref{nonlocal_equations}) we get
(\ref{einstein1}) and (\ref{einstein2}). The fields $\psi(t)$ and $\xi(t)$, defined by (\ref{psiidenti}) and (\ref{xiidenti}) respectively, satisfy (\ref{equpsi}) and (\ref{equxi}).

In the previous section we have obtained power-law solutions for the
local formulation of the original nonlocal gravity, where two
scalars, $\psi$ and $\xi$ have been introduced in the
action~(\ref{action2}). Thus, we come to conclusion that these
solutions are solutions of the initial nonlocal model as well. This
can also be checked immediately by direct substitution.

In the FLRW metric, the d'Alembert operator acting on a scalar $A(t)$ can be expressed as
\begin{equation}
\Box A\equiv \frac1{\sqrt{-g}} \, \partial_{\rho} \left(\sqrt{-g} \,
g^{\rho\sigma} \partial_{\sigma}\right) A= -\frac{1}{a^3}
\frac{d}{dt} \left( a^3 \frac{dA}{dt}\right) \; ,
\end{equation}
while its inverse operator reduces to a double
integration~\cite{Deser:2007jk,Woodard}:
\begin{equation}\label{1_Box}
\Box^{-1}\left[A(t)\right]={}-\int\limits^t_{\tilde{t}_0}\frac{d\tilde{t}}{a^3(\tilde{t})}
\int\limits^{\tilde{t}}_{\eta_0}d\eta a^3(\eta)
A(\eta)\,.
\end{equation}
where $\tilde{t}_0$ and $\eta_0$ are two initial boundaries for the integrals. For the
power-law solution with $H=n/t$, we get $R=6n(2n-1)/t^2$ and the
solutions can correspondingly be obtained by integration, as
\begin{equation*}
\psi(t)={}-6n(2n-1)\int\limits^t_{\tilde{t}_0}\frac{d\tilde{t}}{\tilde{t}^{3n}}
\int\limits^{\tilde{t}}_{\eta_0} \eta^{3n-2}d\eta={}-\frac{6n(2n-1)}{3n-1}\ln\left(\frac{t}{t_0}\right)+\psi_1t^{1-3n}\,,
\end{equation*}
where the integration constants, $t_0$ and $\psi_1$, are  connected
with $\tilde{t}_0$ and $\eta_0$. This solution coincides with
(\ref{psisoln}). Setting $\psi_1=0$, which corresponds to $\tilde{t_0}=t_0$ and $\eta_0=0$, we get
\begin{equation*}
\xi(t)=\frac{(3n-1)mf_0}{t_0^m}\int\limits^t_{t_0}\frac{d\tilde{t}}{\tilde{t}^{3n}}
\int\limits^t_{t_1}\eta^{3n+m-2}d\eta\,,
\end{equation*}
where $m$ is defined in Eq.~(\ref{fm}). Therefore, we
recover the solution~(\ref{xisoln}).
We thus conclude that the
power-law solutions found in the previous sections are equivalent to those corresponding to
the original form of the nonlocal theory.

It should be noted that the initial nonlocal model might be
non-equivalent to its local formulation. Actually this non-equivalence
does not arise from a difference in the equations,
but from the initial (boundary) conditions.
Let us make some further comments on this issue.
By recasting the original form into the biscalar-tensor
representation, one needs to invert the relationship
$\psi=\Box^{-1}R$, in the form: $\Box\psi=R$. For a given
background, the solution for the latter equation is unique up to a
harmonic function $\chi$ which satisfies $\Box\chi=0$, hence causing
a legitimate problem, as reported in
the first paper of Ref. [32].~\footnote{Note that in this paper the non-equivalence between biscalar-tensor representation of
nonlocal gravity and its original form have been shown explicitly in the case of $f=$const only, when the original model is local.}

Compared to the original form~(\ref{action}), the scalar-tensor
presentation seems to have introduced a new degree of freedom
$\chi$, as pointed out in~\cite{Woodard,Koivisto:2010}. In fact, this can be seen more
clearly if we write
\begin{equation*}
\psi\longrightarrow\psi+\chi
\end{equation*}
 into (\ref{action2}), the corresponding term being
\begin{equation}
\xi(\Box\psi-R)\longrightarrow\xi\big(\Box(\psi+\chi)-R\big)\,,
\end{equation}
and, after integration by parts, the change is
\begin{equation}
g^{\mu\nu}\partial_\mu\xi\partial_\nu\psi\longrightarrow
g^{\mu\nu}\partial_\mu\xi\partial_\nu(\psi+\chi)\,.
\end{equation}
Hence, it seems to have added an extra degree of freedom to the
Lagrangian which is absent in the original form. However, if we
impose an appropriate boundary condition, for example, $\chi=0$ to
recover the original form, then this would-be extra degree of
freedom may be eliminated in this way. The issue on the choice of a correct
boundary condition should be the only non-equivalence between the
original form and its biscalar-tensor representation. Thus, for instance,
in~\cite{Woodard} the authors determine the inverse d'Alembert
operator using the retarded Green function, in other words, they
fix a solution of the equation $\Box R=0$ putting $\tilde{t}_0=0$
and $\eta_0=0$.

Power-law solutions display a singularity at $t=0$, so for such
solutions it would be better to choose a positive value of
$\tilde{t}_0=t_0$. In Sect.~\ref{jordan} we obtain that for the
model with nonzero $\Lambda$ the value of $t_0$ is defined by
$\Lambda$, whereas at $\Lambda=0$, $t_0$ is an arbitrary number,
defined by the initial conditions, in particular by $\rho_0$.

A final comment is in order. As stated above, the biscalar-tensor representation introduces two
scalars, $\psi$ and $\xi$, therefore, working in this way it
seems that one will encounter a ghost-like behavior, as pointed out
in Refs.~\cite{Koivisto2,Nojiri:2010pw,Bamba1104,ZS}. However, since
the original nonlocal model does not introduce any new degree of
freedom, the ghost-like behavior of the biscalar-tensor theory may
not be physically relevant at all. Indeed, the associated terms can be cast
as a boundary term of the nonlocal operators~\cite{Koivisto:2010}. At
the classical level, a necessary way to check whether the
ghost-like behavior is physically relevant or not is by considering
the equivalence of the solutions coming from the original nonlocal
formulation and from its biscalar-tensor form, respectively.

\section{Action and equation of motion in the Einstein frame}
\label{JE}
\subsection{The Jordan and Einstein frames}
Once a modified gravity theory is recast into its scalar-tensor
presentation, it immediately follows that both the Jordan frame
(where the matter sector minimally couples to gravity) and the
Einstein one (where the Ricci is linear but matter couples to
gravity non-minimally) are available (for a description
see~\cite{BD:1961}). These two frames are related by conformal
transformation
\begin{equation}
g_{\mu\nu}=\Omega^2 g_{\mu\nu}^{(E)}\,, \label{tildeg}
\end{equation}
where we denote the metric in the Jordan frame by ${g}_{\mu\nu}$,
while the one in the Einstein frame is labeled as
${g}^{(E)}_{\mu\nu}$.

One soon realizes that the conformal transformation connecting both
frames cannot be simply interpreted as a coordinate transformation
of the theory, and this is the reason why there has been a long
debate on which of these two frames is `the physical one',
regardless of the fact that the mathematical equivalence of the two
frames is quite clear~\cite{Faraoni_OLD,Faraoni-Nadeau,CNOT}. Recent
researches have further clarified that, at least at the classical
level, the two frames are physically equivalent, what means that all
observational quantities (e.g., the redshift $z$) should yield the
same value in both cases~\cite{NS:2010a}.

Based on the corresponding solutions in both frames, one can
furthermore probe the equivalence of both frames in the framework of
nonlocal gravity  inspired models. It is of interest to check the
precise behavior of the corresponding solutions in the Einstein
frame to see if and, being the case, how much they differ from those
obtained in the Jordan frame. Moreover, we will show that the transitions
between these frames is a useful tool for the construction of
power-law solutions in the Einstein one.

\subsection{Conformal transformation}
We are investigating the nonlocal model (\ref{action2}) in the
Einstein frame with a perfect fluid. It has been shown that those
fluids have the same EoS in both
frames~\cite{Book-Capozziello-Faraoni}, which implies that the
matter sector remains unaltered in both. Hence, it will be
interesting to trace the behavior of the cosmological solutions in
the Einstein frame, with the same matter fields.

On the other hand, it is known that a theory with higher derivatives
in the action will often suffer from the ghost problem, i.e. a wrong
sign in front of the kinetic terms, which will cause an instability
problem thus making the theory physically irrelevant. Therefore, it
is important to examine if the theory contains a ghost or not. To
see this non-perturbatively, one needs first make a conformal
transformation of the metric to bring the action into the form of
the one in the Einstein frame~\cite{Nojiri:2010pw,Bamba1104,ZS},
namely the conformal frame in which the gravitational part of the
action (\ref{action2}) becomes purely Einsteinian. Note that the
matter field is assumed to be minimally coupled to gravity in the
Jordan frame, as given in action~(\ref{action2}). Power-law
solutions in the Einstein frame for the model given by the action
(\ref{action2}) without matter were considered
in~\cite{Bamba1104,ZS}. In what follows we shall denote all quantities
in the Einstein frame by adding a tag $~^{(E)}$ to the corresponding
ones in the other frame, in order to avoid confusion.

Let us  consider the conformal transformation (\ref{tildeg}).
 Using $g^{\mu\nu}=\Omega^{-2}
{g^{\mu\nu}}^{(E)}$, one obtains the relationship between the Ricci
scalars in the two frames
\begin{equation}
\label{conformtrans} {R}=
\Omega^{-2}\left[R^{(E)}-6\left(\Box^{(E)}\ln{\Omega}
+g^{\mu\nu(E)}\nabla^{(E)}_\mu\ln{\Omega}\nabla^{(E)}_\nu\ln{\Omega}\right)
\right]
\end{equation}
and, inserting this into action (\ref{action2}), one immediately
identifies the conformal factor as~\cite{Nojiri:2010pw,Bamba1104,ZS}
\begin{equation}
 \Omega^{-2}=\Psi\equiv1+f(\psi)+\xi\,, \label{Omega}
\end{equation}
Then, by introducing a new field $\phi$.
\begin{equation}
\label{conformfactor}
\phi\equiv\ln{\Omega}={}-\frac{1}{2}\ln{\left(1+f(\psi)+\xi\right)}={}-\frac{1}{2}\ln(\Psi)\,,
\end{equation}
to remove the Lagrangian multiplier $\xi$, we finally get
the following action in the Einstein frame:
\begin{equation}
\label{finalver}
\begin{split}
S=&\int\!
d^4x\sqrt{-g^{(E)}}\left\{\frac{1}{2\kappa^2}\left[R^{(E)}-6
\nabla^{\mu(E)}\phi\nabla_\mu^{(E)}\phi-2\nabla^{\mu(E)}\phi\nabla_\mu^{(E)}\psi \right.\right.\\
-&\left.\left.e^{2\phi}f_{,\psi}(\psi)\nabla^{\mu(E)}\psi\nabla_\mu^{(E)}\psi-2e^{4\phi}\Lambda\right]
  +e^{4\phi}\mathcal{L}_\mathrm{m}(Q;\, e^{2\phi}g^{(E)})\right\}\,.
\end{split}
\end{equation}
In what follows we will derive the corresponding equations of motion
in the Einstein frame by varying this action. After solving them we
will discuss specific cosmological behaviors.

\subsection{Equations of motion}

By varying action (\ref{finalver}) with respect to the metric
${g^{\mu\nu}}^{(E)}$ one obtains the corresponding Einstein
equations, as follows
\begin{equation}
\label{einEOM}
\begin{split}
R_{\mu\nu}^{(E)}&{}-\frac{1}{2}g_{\mu\nu}^{(E)}\left[R^{(E)}-g^{\alpha\beta(E)}
\left(6\partial_\alpha\phi\partial_\beta\phi
+2\partial_\alpha\phi\partial_\beta\psi+e^{2\phi}f_{,\psi}\partial_\alpha\psi\partial_\beta\psi\right)
-2e^{4\phi}\Lambda
\right]-{}\\&{}-6\partial_\mu\phi\partial_\nu\phi-2\partial_\mu\phi\partial_\nu\psi
-e^{2\phi}f_{,\psi}\partial_\mu\psi\partial_\nu\psi=\kappa^2T_{\mu\nu}^{(E)}\,,
\end{split}
\end{equation}
where we recall that $\phi$ is defined by~(\ref{conformfactor}), and
the energy--momentum tensor in the Einstein frame  as
\begin{eqnarray}\label{TEJ}
T_{\mu\nu}^{(E)}\equiv-\frac{2}{\sqrt{-g^{(E)}}}\frac{\delta\left(\Omega^4\sqrt{-g^{(E)}}
\mathcal{L}_\mathrm{m}\right)}{\delta
g^{\mu\nu(E)}}=e^{2\phi}T_{\mu\nu}\,.
\end{eqnarray}
The field equations read
\begin{eqnarray}
\Box^{(E)}(6\phi+\psi)-e^{2\phi}f_{,\psi}g^{\mu\nu(E)}\partial_\mu\psi\partial_\nu\psi-4e^{4\phi}\Lambda
+\kappa^2e^{4\phi}\left(
4\mathcal{L}_\mathrm{m}+\frac{\partial{\mathcal{L}_\mathrm{m}}}{\partial\phi}\right)&=&0\,,
\label{einscal}\\
2\Box^{(E)}\phi+2e^{2\phi}f_{,\psi}\Box^{(E)}\psi+g^{\mu\nu(E)}e^{2\phi}
\left(4f_{,\psi}\partial_\mu\psi\partial_\nu\phi
   +f_{,\psi\psi}\partial_\mu\psi\partial_\nu\psi\right)
&=&0\,.\label{einsca2}
\end{eqnarray}
At first sight, one may guess that the last term of
Eq.~(\ref{einscal}) could be troublesome. However, this term can be
substituted by a combination of the conformal factor and the trace
part of the energy--momentum tensor in the Einstein frame. To
achieve this, from Eqs.~(\ref{tildeg}) and (\ref{TEJ}), we get
\begin{eqnarray}
T_\mu^{\nu(E)}&=&T_{\mu\alpha}^{(E)}g^{\alpha\nu(E)}=\Omega^2T_{\mu\alpha}
\Omega^2g^{\alpha\nu}=\Omega^4T_{\mu}^\nu\,,\label{TJE}\\
\frac{\partial\mathcal{L}_\mathrm{m}}{\partial\phi}&=&
\frac{\partial\mathcal{L}_\mathrm{m}}{\partial g^{\mu\nu}}
\frac{\partial g^{\mu\nu}}{\partial\phi}=
\frac{\partial\mathcal{L}_\mathrm{m}}{\partial g^{\mu\nu}}
\left[\frac{\partial\left(\Omega^{-2}
g^{\mu\nu(E)}\right)}{\partial\phi}\right]=-2\Omega^{-2}g^{\mu\nu(E)}
\frac{\partial\mathcal{L}_\mathrm{m}}{\partial g^{\mu\nu}}
=-2g^{\mu\nu}\frac{\partial\mathcal{L}_\mathrm{m}}{\partial
g^{\mu\nu}},\,\,\,\label{Tphi}
\end{eqnarray}
while from the definition of the energy--momentum tensor in the
Jordan frame, we obtain
\begin{equation}
\label{TJ}
T^\mu_\mu=g^{\mu\nu}\left(g_{\mu\nu}\mathcal{L}_\mathrm{m}-2\frac{\partial\mathcal{L}_\mathrm{m}}{\partial
g^{\mu\nu}}\right)=4\mathcal{L}_\mathrm{m}-2g^{\mu\nu}\frac{\partial\mathcal{L}_\mathrm{m}}{\partial
g^{\mu\nu}}\,.
\end{equation}
Inserting Eqs.~(\ref{TJE}) and (\ref{Tphi}) into (\ref{TJ}), one
recovers the last term of Eq.~(\ref{einscal}), under the form
\begin{equation}
\label{righthand}
4\mathcal{L}_\mathrm{m}+\frac{\partial\mathcal{L}_\mathrm{m}}{\partial\phi}
=\Omega^{-4}T_\mu^{\mu(E)}\,.
\end{equation}

\subsection{The FLRW metric}

As is well known~\cite{Book-Capozziello-Faraoni}, conformally flat
metrics are mapped into each other. The FLRW metric is conformally
flat, so starting from the FLRW metric in the Jordan frame we obtain
the corresponding FLRW metric in the Einstein one. This leads us to
directly start from a FLRW metric with cosmic time in the Einstein
frame
\begin{equation}
\label{powersol}
ds^2={}-dt^{2}_{E}+a^{2}_{E}(t_{E})\delta_{ij}dx^idx^j\,,
\end{equation}
where in $dt_{E}$ and $a_{E}(t_{E})$ the index $E$ denotes the
corresponding quantities in the Einstein frame. We get
\begin{equation}
\label{attrans} dt_E=\Omega^{-1}dt, \qquad a_E=\Omega^{-1}a\,,
\qquad
H_E\equiv\frac{d\log{a_E}}{dt_E}=\Omega\left(H-\frac{\dot\Omega}{\Omega}\right)
\end{equation}

Under the conformal transformation~(\ref{tildeg}), the
energy--momentum tensor of a perfect fluid transforms
as~\cite{Felice_Tsujikawa,Book-Capozziello-Faraoni}
\begin{eqnarray}\label{Tmunv}
T^{\mu(E)}_\nu=\diag(-\rho_E, P_E, P_E, P_E)=\Omega^4\diag(-\rho, P,
P, P)\,.
\end{eqnarray}
Using this equation together with Eq.~(\ref{attrans}), we obtain
that the continuity (conservation) equation~(\ref{equ_rho}) is
transformed into the following one in the Einstein frame
\begin{eqnarray}\label{conserE}
\rho^\prime_E+3H_E(\rho_E+P_E)=\frac{\Omega^\prime}{\Omega}(\rho_E-3P_E)\,,
\end{eqnarray}
where the prime denotes derivative with respect to the cosmological
time in the Einstein frame, i.e. $^\prime\equiv d/dt_E$.

From (\ref{Tmunv}), we obtain the EoS parameter
$w_{\mathrm{m}}=P_{\mathrm{m}}/\rho_{\mathrm{m}}=P_E/\rho_E$,
therefore, the conservation law (\ref{conserE}) can be rewritten as
\begin{equation}\label{conserEfin}
\rho_E^\prime+\rho_E\left[3H_E(1+w_{\mathrm{m}})+\phi^\prime(3w_{\mathrm{m}}-1)\right]=0\,.
\end{equation}
It should be noted that if matter just reduces to radiation
($w_{\mathrm{m}}=1/3$), then the conservation laws in the Jordan and
Einstein frames coincide.

For the model with a perfect fluid,
Eqs.~(\ref{einEOM})--(\ref{einsca2}) acquire the following form, in
the FLRW metric,
\begin{eqnarray}
3H_E^2-3\phi^{\prime2}-\phi^\prime\psi^\prime-\frac{e^{2\phi}}{2}f_{,\psi}\psi^{\prime2}-e^{4\phi}\Lambda
&=&\kappa^2\rho_E\,,
\label{einEOM-1.2}\\
-6(H_E^\prime+2H_E^2)-6\phi^{\prime2}-2\phi^\prime\psi^\prime-e^{2\phi}f_{,\psi}\psi^{\prime2}
+4e^{4\phi}\Lambda
&=&\kappa^2\rho_E(3w_{\mathrm{m}}-1)\,,\label{einEOM-2.2}\\
6\phi^{\prime\prime}+\psi^{\prime\prime}+3H_E(6\phi^\prime+\psi^\prime)-e^{2\phi
}f_{,\psi}\psi^{\prime2}+4e^{4\phi}\Lambda
&=&\kappa^2\rho_E(3w_{\mathrm{m}}-1)\,,
\label{einscal-1.2}\\
2\phi^{\prime\prime}+6H_E\phi^\prime+2e^{2\phi}f_{,\psi}(\psi^{\prime\prime}+3H_E\psi^\prime)
+4e^{2\phi}f_{,\psi}\phi^{\prime}\psi^{\prime}&=&{}-e^{2\phi}f_{,\psi\psi}\psi^{\prime2}\,,\label{einsca2-1.2}
\end{eqnarray}
where the first two are independent Einsteinian equations, while the
other two are scalar field equations. Thus, the complete set of
equations is given by (\ref{einEOM-1.2})--(\ref{einsca2-1.2}).
Eq.~(\ref{conserEfin}) follows from this system. Combining
Eqs.~(\ref{einEOM-1.2}) and (\ref{einEOM-2.2}), one obtains the
remarkably useful result that the equation for a scalar $\phi$ can
be reduced to an algebraic equation, as follows,
\begin{equation}\label{einEOMspe}
\Lambda e^{4\phi}=
H_E^\prime+3H_E^2+\frac{\kappa^2\rho_E}{2}(w_{\mathrm{m}}-1)\,.
\end{equation}
For $\Lambda\neq 0$, one can formally obtain from (\ref{einEOMspe})
the expression for the conformal factor $\phi(t_E)$
\begin{eqnarray}\label{phisol}
\phi(t_E)=\frac{1}{4}\ln\left[\frac{1}{\Lambda}\left(H_E^\prime+3H_E^2+
\frac{\kappa^2\rho_E}{2}(w_{\mathrm{m}}-1)\right)\right]\,.
\end{eqnarray}
It should also be noted that, by combining Eqs.~(\ref{einEOM-2.2})
and (\ref{einscal-1.2}), one can eliminate the $f(\psi)$ and matter
terms to obtain a second-order differential equation in terms of
$\psi(t_E)$ and $\phi(t_E)$, namely
\begin{equation}\label{equsual}
6\phi^{\prime\prime}+\psi^{\prime\prime}+3H_E(6\phi^\prime+\psi^\prime)
+6(H_E^\prime+2H_E^2)+6\phi^{\prime2}+2\phi^\prime\psi^\prime=0\,.
\end{equation}
It is most convenient to derive the expression for $\psi(t_E)$ from
this equation once $H_E$ and $\phi(t_E)$ are known.

\section{Power-law solutions in the Einstein frame for the model without matter}
\label{Einvacuum}

We here investigate power-law solutions (with $H_E=n_E/t_E$) in the
Einstein frame.  We consider the case of an exponential $f(\psi)$,
given by (\ref{f}). For the model with $\rho_E=0$, the cases with
and without matter will turn out to be essentially different.

If $\Lambda=0$, from (\ref{einEOMspe}) we obtain the following
power-law solution
\begin{equation}
\label{H_E_noLambda_nomatter} H_E^\prime=-3H_E^2,\qquad
\Rightarrow\qquad H_E=\frac1{3(t_E-T_0)}\,,
\end{equation}
meaning that all solutions correspond to the power-law behavior of
the Hubble parameter with $n_E=1/3$. Thus, the general solution of
the system (\ref{einEOM-1.2})--(\ref{einsca2-1.2}), which
corresponds to arbitrary initial conditions for the scalar fields,
yields just one cosmological evolution, specified
by~(\ref{H_E_noLambda_nomatter}).

In the case where $\Lambda\neq 0$ and matter is absent, from
(\ref{phisol}) we obtain
\begin{eqnarray}\label{phisol1}
\phi(t_E) =\frac{1}{4}\ln{\left[\frac{n_E(3n_E-1)}{\Lambda
t_E^2}\right]}\,.
\end{eqnarray}
Note that there is no solution with  $n_E=1/3$ if $\Lambda\neq 0$.
Eq.~(\ref{equsual}) is a linear differential equation for
$\psi(t_E)$. Inserting the function $\phi(t_E)$ we just obtained
into this equation, we get the following solution
\begin{eqnarray}\label{psisol1}
\psi(t_E)
=\widetilde{\psi}_0t_E^{2-3n_E}+m_E\ln{\left(\frac{t_E}{\widetilde{t}_0}\right)}\,,
\end{eqnarray}
where $m_E\equiv3(2n_E-1)(4n_E-3)/[2(2-3n_E)]$, while
$\widetilde{\psi}_0$ and $\widetilde{t}_0$ are two integration
constants. Inserting (\ref{phisol1}) and (\ref{psisol1}) into the
system~(\ref{einEOM-1.2})--(\ref{einsca2-1.2}), one gets the
following constraints on the integration constants
\begin{eqnarray}\label{psisol1a}
\widetilde{\psi}_0&=&0\,,\qquad \widetilde{t}_0=3f_0(2n_E-1)\sqrt{\frac{n_E}{\Lambda(3n_E-1)}}\,,\\
\alpha m_E&=&1 \Longleftrightarrow
\alpha+\frac{2(3n_E-2)}{3(2n_E-1)(4n_E-3)}=0\,,
\end{eqnarray}
from which the two branches corresponding to the index $n_E$ are
expressed in terms of the parameter $\alpha$, as
\begin{equation} \label{neinsteinvacuum}
\left\{ \begin{aligned}
         n_{E(1)} &= \frac{15\alpha-3+\sqrt{3(3\alpha^2+2\alpha+3)}}{24\alpha}\,, \\
                  n_{E(2)} &=
                  \frac{15\alpha-3-\sqrt{3(3\alpha^2+2\alpha+3)}}{24\alpha}\,.
\end{aligned} \right.
\end{equation}
We note that this solution is the same as the one found
in~\cite{ZS}, where the vacuum solution with non-vanishing
cosmological constant $\Lambda$ is constructed by conformal
transformation from the corresponding one in the Jordan frame. Also
we note that for any range of the parameter $\alpha$, the range for
$n_{E(1)}$ is  $n_{E(1)}\in(1/2,~3/4)$, while for $n_{E(2)}$ we have
$n_{E(2)}\in(-\infty,~1/2)\cup(3/4,~+\infty)$.

There are a few special cases of the parameter $n_E$ for which the
above mentioned solution does not exist. In these cases the
parameter $m_E$ is either equal to zero or it does not exist. Let us
consider all of them in detail.
\begin{itemize}
\item If $n_E=1/2$, then
\begin{equation}
\label{psi_1_2} \psi(t_E) = c_1+ c_2\sqrt{t_E}.
\end{equation}
Substituting this solution into Eq.~(\ref{einEOM-1.2}), we conclude
that this equation is not satisfied whatever be the constants $c_1$
and $c_2$.

\item If $n_E=2/3$, then
\begin{equation}
\label{psi_2_3} \psi(t_E) =
\frac{1}{12}\ln\left(\frac{t_E}{\widetilde{T}}\right)^2+c_1\ln\left(\frac{t_E}{\widetilde{T}}\right).
\end{equation}
Substituting this solution into Eq.~(\ref{einEOM-1.2}), we conclude
that this equation is not satisfied for any value of the constants
$c_1$ and $\widetilde{T}$.

 \item If $n_E=3/4$, then
\begin{equation}
\label{psi_3_4} \psi(t_E) = c_1+ c_2t_E^{-1/4}.
\end{equation}
Substituting this solution into Eq.~(\ref{einEOM-1.2}), with the
condition $c_2=0$, from Eq.~(\ref{einsca2-1.2}) we conclude that
this equation is not satisfied for any value of  $c_1$.
\end{itemize}

In summary, we have obtained power-law solutions in the Einstein
frame for the model without matter, as follows,
\begin{eqnarray}
H(t_E)&=&\frac{n_E}{t_E}\,,\\
\phi(t_E)&=&\frac{1}{4}\ln{\left[\frac{n_E(3n_E-1)}{\Lambda
t_E^2}\right]}\,,\label{phisolution}\\
\psi(t_E)&=&\frac{1}{\alpha}\ln{\left[\frac{t_E}{3f_0(2n_E-1)}
\sqrt{\frac{\Lambda(3n_E-1)}{n_E}}\right]}\label{psisolution}\,,
\end{eqnarray}
where the parameter $n_E$ is connected with $\alpha$ by
$(\ref{neinsteinvacuum})$. Note that $n_E\neq 1/2$, $n_E\neq 2/3$,
$n_E\neq 3/4$, and $n_E\neq 1/3$. This is valid for $\Lambda\neq 0$.
If $\Lambda=0$, then all solutions of this system have power-law
behavior with $n_E= 1/3$ and observe that the solution obtained has no
arbitrary integration constant.

\section{Power-law solutions in the Einstein frame for the model with matter}
\label{einsolutions}

\subsection{The case $\Lambda=0$}
\label{lambdacase}

For $H={n_E}/{t_E}$, from Eq.~(\ref{einEOMspe}) one gets that
\begin{eqnarray}\label{rhoein1}
\rho_E(t_E)=\frac{2n_E(3n_E-1)}{\kappa^2(1-w_\mathrm{m})t_E^2}\,.
\end{eqnarray}
Inserting this into Eq.~(\ref{conserEfin}), the expression for
$\phi(t_E)$ follows
\begin{eqnarray}\label{rhos1}
\phi(t_E)=\frac{2-3n_E(1+w_\mathrm{m})}{3w_\mathrm{m}-1}\ln\left(\frac{t_E}{\widetilde{t}_1}\right)\,,
\end{eqnarray}
where $\widetilde{t}_1$ is an integration constant.
Formula~(\ref{rhos1}) is valid for  $n_E\neq1/3$,
$w_\mathrm{m}\neq1/3$, and $w_\mathrm{m}\neq1$. Furthermore, using
Eq.~(\ref{equsual}) we obtain the solution for $\psi(t_E)$:
\begin{eqnarray}\label{psis1}
\psi(t_E)=\widetilde{\psi}_1\left(\frac{t_E}{\widetilde{t}_2}\right)^{1+\frac{4+3n_E(w_\mathrm{m}-3)}{1-3w_\mathrm{m}}}+\frac{
12(2n_E-1)\left[3(1-w_\mathrm{m})+n_E(3w_\mathrm{m}-5)\right]}
{\left[5-3w_\mathrm{m}+3n_E(w_\mathrm{m}-3)\right](3w_\mathrm{m}-1)}
\ln\!\left(\frac{t_E}{\widetilde{t}_2}\right),
\end{eqnarray}
where $\widetilde{t}_2$ and $\widetilde{\psi}_1$ are integration
constants. Inserting Eqs.~(\ref{rhoein1}), (\ref{rhos1}), and
(\ref{psis1}) into the
system~(\ref{einEOM-1.2})--(\ref{einsca2-1.2}), one obtains the
following constraints on the integration constants and parameters
\begin{eqnarray}
\widetilde{\psi}_1&=&0\,,\label{constr1-1}\\
n_{E\pm}&=&\frac{3\left[(3w_\mathrm{m}^2-11)+2\alpha(11-9w_\mathrm{m})\right]\pm|1-3w_\mathrm{m}|\sqrt{3\left[3(2\alpha-w_\mathrm{m}+1)^2-16\alpha\right]}}
{6\left[3(w_\mathrm{m}-1)^2+4\alpha(5-3w_\mathrm{m})-12\right]}\,,\label{constr1-2}\\
\frac{\widetilde{t}_2}{\widetilde{t}_1}&=&\left[\frac{(3n_E-1)(3w_\mathrm{m}-1)}{6(2n_E-1)(w_\mathrm{m}-1)f_0}\right]^{\frac{3w_\mathrm{m}-1}
{2\left[2-3n_E(1+w_\mathrm{m})\right]}}\label{constr1-3}\,.
\end{eqnarray}
It follows that Eq.~(\ref{constr1-2}) gives rise to constraints on
the parameters $\alpha$ and $w_\mathrm{m}$, for $n_{E\pm}$ to be a
real number:
\begin{eqnarray}
3(2\alpha-w_\mathrm{m}+1)^2-16\alpha\geqslant0\,.\label{alconstr}
\end{eqnarray}

Interestingly enough, in the limit $\alpha\gg1$, using
Eq.~(\ref{constr1-2}), we obtain the following asymptotic behavior
for the index $n_{E\pm}$
\begin{eqnarray}
n_{E\pm}(\alpha\gg1)\approx\frac{11-9w_\mathrm{m}\pm|1-3w_\mathrm{m}|}
{4(5-3w_\mathrm{m})}\,.\label{nayp}
\end{eqnarray}
One immediately realizes that, for $w_\mathrm{m}=-1$,
$n_{E+}\longrightarrow0.75$ asymptotically, what corresponds to an
upper bound on the value for the index in the vacuum case, in the
Einstein frame (see Eq.~(38) of Ref.~\cite{ZS}). In fact, when
$w_\mathrm{m}=-1$, the solutions (\ref{rhoein1})--(\ref{constr1-3})
reduce to the following form:
\begin{eqnarray}\label{0lamspe}
\rho_E(t_E)=\frac{n_E(3n_E-1)}{\kappa^2t_E^2}\,,\quad\phi(t_E)
=-\frac{1}{2}\ln\left(\frac{t_E}{\widetilde{t}_1}\right)\,,\quad
\psi(t_E)=\frac{1}{\alpha}\ln\left[\frac{3n_E-1}{3f_0\widetilde{t}_1(2n_E-1)}\right],
\end{eqnarray}
while the corresponding energy density in the Jordan frame is just
the cosmological constant $\Lambda$. By using conformal
transformation Eq.~(\ref{Tmunv}) and comparing it with the solution
for $\rho_E(t_E)$ in (\ref{0lamspe}), we can express the integration
constant $\widetilde{t}_1$ in terms of $\Lambda$, as
\begin{eqnarray}\label{t1special}
\widetilde{t}_1=\sqrt{\frac{n_E(3n_E-1)}{\kappa^2\Lambda}}\,.
\end{eqnarray}
It is straightforward to see that Eq.~(\ref{0lamspe}) recovers the
vacuum solutions, Eqs.~(\ref{phisolution})--(\ref{psisolution}).

Another interesting asymptotic behavior is the one for
$w_\mathrm{m}>1/3$, $n_{E+}\longrightarrow1/2$, while for
$w_\mathrm{m}<1/3$ we get $n_{E-}\longrightarrow1/2$, regardless of
the value of the EoS parameter $w_\mathrm{m}$.

We thus have derived solutions for all nonzero values of $n_E$, but
for $n_E=1/3$. Also, we assume that $w_\mathrm{m}\neq 1/3$ and
$w_\mathrm{m}\neq 1$. Solutions in these special cases will be
considered in the next section, by conformally transforming the
power-law solutions obtained in the Jordan frame into the Einstein
one.

\subsection{The case  $\Lambda\neq0$}
\subsubsection{$w_\mathrm{m}= 1$}
In this case, we cannot obtain a general solution for the system
(\ref{einEOM-1.2})--(\ref{einsca2-1.2}). Nevertheless, from
Eq.~(\ref{phisol}) one finds that $w_\mathrm{m}=1$ is a special case
where the function $\phi(t_E)$ has already been found.
Thus, in the following we will consider the case $w_\mathrm{m}= 1$.

From Eq.~(\ref{phisol}) one obtains
\begin{equation}\label{philam1}
\phi(t_E)=\frac{1}{4}\ln\left[\frac{(3n_E-1)n_E}{\Lambda
t_E^2}\right],
\end{equation}
and the condition $n_E\neq 1/3$. Substituting this function into
Eq.~(\ref{conserEfin}), we get
\begin{equation}
\rho_E(t_E)\propto t_E^{1-6n_E}.
\end{equation}
On the other hand, inserting the expression (\ref{philam1}) into
Eq.~(\ref{equsual}), $\psi(t_E)$ can be solved as
\begin{equation}
\psi(t_E)={}-\frac{3(2n_E-1)(4n_E-3)}{2(3n_E-2)}\ln\left(\frac{t_E}{T_1}\right)+C_1t_E^{2-3n_E},
\end{equation}
where $C_1$ and $T_1$ are arbitrary constants and $n_E\neq 2/3$,
$n_E\neq 1/2$, $n_E\neq 3/4$.

However, substituting this function into (\ref{einEOM-1.2}), we
discover that this equation cannot hold for any values of $C_1$ and
$T_1$. Note that for $n_E=1/2$, $n_E= 2/3$, and $n_E= 3/4$,
Eq.~(\ref{equsual}) has different solutions\footnote{These solutions
coincide with (\ref{psi_1_2})--(\ref{psi_3_4}).},  but the system of
equations (\ref{einEOM-1.2})--(\ref{einsca2-1.2}) is not satisfied
for these values of $n_E$, either. Thus, we conclude that there is
no consistent solution with $w_\mathrm{m}=1$.

\subsubsection{$w_\mathrm{m}\neq 1$}
If $w_\mathrm{m}\neq 1$, then we can express $\phi(t_E)$ via
$\rho_E(t_E)$ using Eq.~(\ref{phisol})  and, furthermore, obtain the
following master equation for $\rho_E(t_E)$ from
Eq.~(\ref{conserEfin}):
\begin{eqnarray}\label{rhon}
\frac{A_1(\rho_E)+A_2(\rho_E)}
{4t_E\left[\kappa^2\rho_Et_E^2(w_\mathrm{m}-1)+2n_E(3n_E-1)\right]}=0,
\end{eqnarray}
where, to simplify the expression, we have defined $A_1(\rho_E)$ and
$A_2(\rho_E)$ as the following two functions of $\rho_E(t_E)$
\begin{eqnarray}
A_1(\rho_E)&\equiv&4n_E\rho_E\left[3\kappa^2(w_\mathrm{m}^2-1)
\rho_Et_E^2+(3n_E-1)\left(6n_E(1+w_\mathrm{m})+1-3w_\mathrm{m}\right)\right]\,,\\
A_2(\rho_E)&\equiv&\rho_E^\prime
t_E\left[3\kappa^2(w_\mathrm{m}^2-1)\rho_Et_E^2+8n_E(3n_E-1)\right]\,.
\end{eqnarray}
For generic values of the parameters $w_\mathrm{m}$ and $n_E$ the
general solution cannot be found in terms of elementary functions
and it is only possible to cast this equation in transcendental
form. At the same time, it is actually easy to solve it for some
particular values of the constants, as the ones which follow.

\begin{itemize}
\item For $n_E=1/3$, we obtain,  from
Eq.~(\ref{phisol}),
\begin{equation}
\label{rhophi13}
    \rho_E(t_E) = \frac{2\Lambda e^{4\phi(t_E)}}{(w_\mathrm{m}-1)\kappa^2},
\end{equation}
while the master equation Eq.~(\ref{rhon}) has the general solution
\begin{equation}
\rho_E(t_E) = \widetilde{\rho}_0 t_E^{-4/3}.
\end{equation}
A straightforward calculation shows that the system
(\ref{einEOM-1.2})--(\ref{einsca2-1.2}) has no solution for
$n_E=1/3$ if $w_\mathrm{m}\neq \pm 1$. (Note that we always assume
that $w_\mathrm{m}\neq -1$, because matter with $w_\mathrm{m}= -1$
coincides with the cosmological constant.)

\item $n_E=1/2$\\
\\
In this case, solving the
system~(\ref{einEOM-1.2})--(\ref{einsca2-1.2}), it is found that
there only exists the trivial solution $\psi_E(t_E)=const$.
Recalling the action (\ref{action}) we are considering, such a
constant solution just corresponds to a rescaling of the Ricci
scalar $R$, or equivalently, to a rescaling of both the Newtonian
constant $G\equiv8\pi/\kappa^2$ and the cosmological
constant~$\Lambda$.
\end{itemize}

The correspondence of the special solutions considered above is that,
in the Einstein frame, the case of a nonvanishing cosmological constant
does not admit a power-law behaved matter sector other than
$\rho_E\propto t_E^{-2}$ for the scaling solution $H_E\propto
n_E/t_E$. To see this clearly, we consider the ansatz
$\rho_E(t_E)=\widetilde{\rho}_1 t_E^{p}$. Then the master equation
for the matter sector, Eq.~(\ref{rhon}), reduces to the following
form
\begin{eqnarray}\label{rhosol0}
\frac{4n_E(3n_E-1)\left[1+2p-3w_\mathrm{m}+6n_E(1+w_\mathrm{m})\right]+
3\widetilde{\rho}_1\kappa^2t_E^{2+p}(4n_E+p)(w_\mathrm{m}^2-1)}
{4t_E\left[\kappa^2\widetilde{\rho}_1t_E^{p+2}(w_\mathrm{m}-1)+2n_E(3n_E-1)\right]}=0\,.
\end{eqnarray}

Let us first consider the case $p\neq -2$. Eq.~(\ref{rhosol0}) is
then satisfied if and only if each of the two terms in the numerator
are equal to zero. For the first term  to vanish it is required
that\footnote{We always assume that $n_E\neq 0$.} $n_E=1/3$ or
$1+2p-3w_\mathrm{m}+6n_E(1+w_\mathrm{m})=0$, while for the second,
that $p=-4n_E$ (recall that $w_\mathrm{m}=\pm1$ is excluded here),
from where the only allowed possibility here is $n_E=1/3$, $p=-4/3$.
As discussed above, this does not satisfy the EOM
(\ref{einEOM-1.2})--(\ref{einsca2-1.2}).

Solutions with $p=-2$ correspond to the ones will be considered in
Sect.~\ref{rhoradiation}.

\section{Relationship between power-law solutions in the Jordan and in the Einstein frames}
\label{conform}
\subsection{Conformal transformation between power-law solutions}
\label{CT}
\subsubsection{General expression for the conformal factor $\Omega$}
In Sects.~\ref{Einvacuum} and~\ref{einsolutions} we have found
solutions for the matter and massless sectors respectively. However, we
recall that in Sect.~\ref{lambdacase}, general solutions were found
except for some singular values of $n_E$ and $w_\mathrm{m}$, since
it was not possible to separate variables in the system
(\ref{einEOM-1.2})--(\ref{einsca2-1.2}) to explicitly solve for the
functions $\psi$ and $\phi$ in these special cases.

Meanwhile, in the vacuum case, by using conformal transformation~\cite{ZS}
it has been found that some power-law solutions in the Jordan frame
correspond to other power-law ones in the Einstein frame.
Here we will generalize this correspondence for the model with a
matter sector, and furthermore use the conformal transformation to
obtain those special solutions in the Einstein frame which are very
difficult to obtain by directly solving the system
(\ref{einEOM-1.2})--(\ref{einsca2-1.2}).

First, we formulate the differential equation for the conformal
factor under which power-law solutions in the Jordan frame
correspond to other power-law solutions in the Einstein frame. Using
(\ref{attrans}), we have
\begin{equation}
H_E=\frac{n_E}{t_E}=\frac{n}{t}\Omega(t)-\dot\Omega(t)\,,
\end{equation}
where we recall that $\dot~\equiv d/dt$. This immediately gives a
relationship between the cosmological times in the Jordan frame,
$t$, and in the Einstein one, $t_E$, as follows
\begin{equation}
t_E=\frac{n_Et}{n\Omega(t)-\dot\Omega(t)t}.
\end{equation}
Taking the derivative of the equation above with respective to $t$
and inserting $d t_E/dt=\Omega^{-1}$, we obtain the following
differential equation for the conformal factor $\Omega(t)$:
\begin{equation} \label{equOmega} \ddot
\Omega-\frac{1}{n_E\Omega}\dot\Omega^2+\frac{n(2-n_E)}{n_E t}\dot
\Omega+\frac{n(n_E-n)}{n_Et^2}\Omega=0.
\end{equation}
This equation has the following general solution:
\begin{equation}
\label{Omega_t} \Omega(t)=\left\{
\begin{array}{lll}
\left(\frac{t}{T}\right)^{n}\left[B_0\left(\frac{t}{T}\right)^{1-n}
+B_1\right]^{\frac{n_E}{n_E-1}},&\qquad n\neq 1,&\quad n_E\neq 1,\\
\left(\frac{t}{T}\right) \left[B_1\ln\left(\frac{t}{T}\right)+1\right]^{\frac{n_E}{n_E-1}},&\qquad n= 1,&\quad n_E\neq 1,\\
B_1t^n\exp\left[\left(\frac{t}{T}\right)^{1-n}\right],&\qquad
n\neq1,&\quad n_E=1,\\
\left(\frac{t}{T}\right)^{B_1}, &\qquad n=1,&\quad n_E=1,
\end{array}
\right.
\end{equation}
where $B_0$, $B_1$ and $T$ are integration constants\footnote{Note
that we use $B_0$ only to include the case $B_0=0$. A nonzero $B_0$ can
always be put equal to one.}.

In principle, as it was done in Ref.~\cite{ZS}, one can now obtain
power-law solutions, by using Eq.~(\ref{Omega_t}), to conformally
transform the corresponding solutions from the Jordan frame to the
Einstein one. However, difficulties arise if one wants to directly
obtain a general solution as~(\ref{Omega_t}).

\subsubsection{Conformal factor corresponding to power-law solutions in the Jordan frame}

As already stated above, although we have obtained the general
solution~(\ref{Omega_t}), the complication of this expression
renders the analysis of the solutions rather difficult. Another
approach to construct the conformal factor $\Omega$ is to insert
solutions found in the Jordan frame into Eq.~(\ref{Omega}), to obtain
the expression directly, case by case. Let us consider the case
$m\neq1-3n$, i.e. use the corresponding solutions (\ref{psisoln})
and (\ref{xisoln}):
\begin{equation}\label{solJordan}
\xi(t)=\xi_0+\xi_1\left(\frac{t}{t_0}\right)^{1-3n}
+\frac{(3n-1)f_0}{3n+m-1}\left(\frac{t}{t_0}\right)^{m},\qquad
\psi(t)=\frac{m}{\alpha}\ln\left(\frac{t}{t_0}\right).
\end{equation}
Thus, one obtains the expression for $\Omega$ by employing
Eq.~(\ref{Omega})
\begin{equation}\label{omegageneral}
\Omega^{-2}(t)=C_1t^{m}+C_2t^{1-3n}+1+\xi_0\quad\Longrightarrow\quad
\Omega(t)=\frac{1}{\sqrt{C_1t^{m}+C_2t^{1-3n}+1+\xi_0}}\,,
\end{equation}
where, for simplicity of the expression, we have defined the two
parameters $C_1$ and $C_2$ as follows
\begin{equation}
\label{define} C_1\equiv \frac{6n+m-2}{3n+m-1}f_0t_0^{-m}\,,\qquad
C_2\equiv\xi_1t_0^{3n-1}\,,
\end{equation}
and again we recall that $m$ is given by (\ref{fm}). Observing
Eq.~(\ref{omegageneral}), we see we need the constraint on the
integration constant $\xi_0=-1$, since otherwise we can never find a
correspondence of power-law solutions between the two frames.
Recalling now Eqs.~(\ref{constantvalues}) and (\ref{nonvanishlamb}),
one immediately realizes the implication of this constraint: there
is no correspondence between power-law solutions in the case of
non-vanishing cosmological constant $\Lambda$ and the matter
sectors.

It is easy to see that there are two simple cases, namely $C_1=0$
and $C_2=0$ where, after conformal transformation, the power-law
Hubble parameter in the Jordan frame $H$ yields a power-law function in
the Einstein frame. We consider these cases separately in the
following subsections. In Sects.~\ref{c10case}--\ref{rhoradiation} we will consider the
case $\Lambda=0$, whereas the case of nonzero $\Lambda$ will be
dealt with in Sect.~\ref{nonzerolam}.

\subsection{The case $C_1=0$}
\label{c10case}
The case $C_1=0$ is special and corresponds to fixing the index $n$.
Indeed, from $C_1=0$, Eq.~(\ref{define}) yields
\begin{equation}\label{finalver1}
n=n_0=\frac{-6+3\alpha\pm\sqrt{3\alpha(3\alpha-4)}}{6(2\alpha-3)}\,.
\end{equation}
 Comparing (\ref{nbyw}) with (\ref{finalver1}), one sees that
$w_\mathrm{m}=1$. Inserting Eq.~(\ref{omegageneral})
into~(\ref{attrans}), we obtain
\begin{equation}\label{t_Omega}
t=\left[\frac{3(1-n_0)}{2\sqrt{C_2}}t_E\right]^{\frac{2}{3(1-n_0)}},\qquad
\Omega^{-1}=\sqrt{C_2}t^{\frac{1-3n_0}{2}}=\sqrt{C_2}
\left[\frac{3(1-n_0)}{2\sqrt{C_2}}t_E\right]^{\frac{1-3n_0}{3(1-n_0)}}\,,
\end{equation}
 hence, using the relationship (\ref{attrans}), we find $n_E=1/3$, thus the Hubble parameter in
the Einstein frame is
\begin{equation}
\label{HEnE13} H_E(t_E)=\frac{1}{3t_E}\,.
\end{equation}
Using the definition $\phi\equiv\ln{\Omega}$, we obtain the
expression
\begin{equation}\label{c2_phi}
\phi(t_E)=\frac{3n_0-1}{3(1-n_0)}\ln\left(\frac{t_E}{\widetilde{t}_3}\right)\,,
\qquad\widetilde{t}_3\equiv\frac{2C_2^{\frac{1}{3n_0-1}}}{3(1-n_0)}\,.
\end{equation}
Inserting (\ref{HEnE13}) and (\ref{c2_phi}) into
Eq.~(\ref{equsual}), $\psi(t_E)$ is found to be
\begin{equation}\label{Einsola1}
\psi(t_E)=\widetilde{\psi}_4\left(\frac{t_E}{\widetilde{t}_*}\right)^{2+\frac{4}{3(n_0-1)}}
+\frac{4n_0(2n_0-1)}{(n_0-1)(3n_0-1)}
\ln\left(\frac{t_E}{\widetilde{t}_*}\right)\,,
\end{equation}
with two integration constants $\widetilde{\psi}_4$ and
$\widetilde{t}_*$. On the other hand, from (\ref{solJordan}) and
(\ref{t_Omega}), we can rewrite the solution (\ref{psisoln}) in
terms of the variables in the Einstein frame:\footnote{We
recall that the integration constant $\psi_1=0$.}
\begin{equation}\label{Einsola2}
\psi(t_E)=\frac{m}{\alpha}\ln\left(\frac{1}{t_0}
\left[\frac{3(1-n_0)}{2\sqrt{C_2}}t_E\right]^{\frac{2}{3(1-n_0)}}\right)=
\frac{4n_0(2n_0-1)}{(n_0-1)(3n_0-1)}
\ln\left(\frac{t_E}{\widetilde{t}_4}\right),
\end{equation}
where we define the constant $\widetilde{t}_4$ as follows:
\begin{equation}\label{t4}
\widetilde{t}_4=\frac{2\sqrt{C_2}}{3\left(1-n_0\right)}t_0^{\frac{3(1-n_0)}{2}}\,.
\end{equation}
Comparing Eqs.~(\ref{Einsola1}) and (\ref{Einsola2}), one
immediately finds that $\widetilde{\psi}_4=0$ and
$\widetilde{t}_*=\widetilde{t}_4$. Thus, we have obtained a new
solution for $n_E=1/3$ and $w_\mathrm{m}=1$ in the Einstein frame by using
conformal transformation, which would had been very difficult to obtain
directly from the system (\ref{einEOM-1.2})--(\ref{einsca2-1.2}).

To summarize, by using conformal transformation, we have found the
following, new solution in the Einstein frame:
\begin{equation} \label{solEin}
\left\{ \begin{aligned}
H_E(t_E)&=\frac{1}{3t_E},\\
\phi(t_E)&=\frac{3n_0-1}{3(1-n_0)}\ln\left(\frac{t_E}{\widetilde{t}_3}\right), \\
\psi(t_E)&=\frac{4n_0(2n_0-1)}{(n_0-1)(3n_0-1)}
\ln\left(\frac{t_E}{\widetilde{t}_4}\right),\\
\rho(t_E)&=\widetilde{\rho}_3 t_E^{\frac{4}{3(n_0-1)}},\\
                          \end{aligned} \right.
                          \end{equation}
with EoS $w_\mathrm{m}=1$ and a constant $\widetilde{\rho}_3$
constrained by (\ref{einEOM-1.2})--(\ref{einsca2-1.2}), as follows
\begin{equation}
\widetilde{\rho}_3=\frac {8{n_0}\,{f_0}\, \left( 1-2n_0 \right) }{3
\kappa^2 \left( 1-{n_0}\right)^2
}\left(\frac{\widetilde{t}_3}{\widetilde{t}_4^{\,2}}\right)^{\frac{2(3n_0-1)}{3(n_0-1)}}\,,\label{constre2}
\end{equation}
where $\widetilde{t}_3$ and $\widetilde{t}_4$ are constants defined
in Eqs.~(\ref{c2_phi}) and (\ref{t4}), respectively.
 Note that $n_0$ is not a free parameter, because it is connected with $\alpha$.
Solutions have been found for arbitrary nonzero $\alpha$, except
for\footnote{The values of $n_0$ equal to $n_0=1/3$, which
corresponds to $\alpha=0$, and $n_0=1/2$, which does not correspond
to any finite value of $\alpha$, are also excluded (see Sect.~3).}
$\alpha=4/3$, which corresponds to $n_0=1$.

\subsection{The case $C_2=0$}
\label{C20case}
In the case $C_2=0$ (or, equivalently,
 $\xi_1=0$), similarly as in
Sect.~\ref{c10case}, the relationship between $t$ and $t_E$ can be
obtained as
\begin{equation}
\label{timetrans}
t=\left(\frac{m+2}{2\sqrt{C_1}}t_E\right)^\frac{2}{m+2}\,.
\end{equation}
Using this relation, the conformal factor can be expressed in terms
of the variables in the Einstein frame, namely
\begin{eqnarray}\label{conformalfact}
\Omega^{-2}(t_E)=C_1\left(\frac{m+2}{2\sqrt{C_1}}t_E\right)^\frac{2m}{m+2}\,.
\end{eqnarray}
Inserting~(\ref{conformalfact}) into Eq.~(\ref{attrans}), we obtain
the Hubble parameter in the
 Einstein frame
\begin{equation}\label{H_E_explicit}
H_E(t_E)=\frac{m+2n}{(m+2)t_E}.
\end{equation}
Thus, by setting the integration constant $\xi_1=0$, we obtain the
correspondence between the power-law solutions in the Jordan and
Einstein frames, and identify the index in the last with
the corresponding one in the Jordan frame, as follows
\begin{equation}
\label{nEin} n_E=\frac{m+2n}{m+2}=
\frac{[3(2\alpha-1)n+1-3\alpha]n}{6\alpha n^2-3n(1+\alpha)+1}\,.
\end{equation}
We recall that the parameter $n$ is determined by $\alpha$ and
$w_\mathrm{m}$ in the constraint (\ref{nbyw}). For practical use, we
rewrite this constraint as
\begin{equation}
\alpha = \frac{(3n-2+3w_\mathrm{m} n)(3n-1)}{6n(2n-1)},
\end{equation}
and substitute it into Eq.~(\ref{nEin}) to express $n$ in terms of $n_E$
and $w_\mathrm{m}$:
\begin{equation}
\label{nin2}
n=\frac{2(2n_E-1)}{3n_E(w_\mathrm{m}+1)-1-3w_\mathrm{m}}\,.
\end{equation}
Inserting (\ref{nin2}) into Eq.~(\ref{conformalfact}), we can also
obtain the expression for the scalar field $\phi$
\begin{equation}\label{conformalein}
\phi(t_E)\equiv\ln\Omega=\frac{2-3n_E(1+w_\mathrm{m})}{3w_\mathrm{m}-1}\ln\left(\frac{t_E}{\widetilde{t}_5}\right),
\end{equation}
where $\widetilde{t}_5$ is an integral constant defined by
\begin{equation}
\widetilde{t}_5=\frac{2}{(m+2)C_1^{\frac{1}{m}}}\,,
\end{equation}
hence connected with $t_0$ and $f_0$ by Eq.~(\ref{define}). We see
that the expression (\ref{conformalein}) coincides with
(\ref{rhos1}) and thus, generally speaking, we get only solutions
previously obtained in Sect.~6.1. Nevertheless, for some values of
the parameters we do not obtain the solutions of Sect.~6.1. Let us
check the possibility to get these solutions using conformal
transformation.

Recall that in Sect.~6.1 we did not find solutions for
$w_\mathrm{m}=1/3$, $w_\mathrm{m}=1$, and $n_E=1/3$.  If
$w_\mathrm{m}=1$, then from Eq.~(\ref{einEOMspe}) it follows that
$n_E=1/3$ and vice versa. Thus, substituting $w_\mathrm{m}=1$ into
(\ref{conformalein}), we get $\phi(t_E)=0$. Straightforward
substitution into Eqs.~(\ref{einEOM-1.2})--(\ref{einsca2-1.2}) shows
that there is no solution in this case.

For $w_\mathrm{m}=1/3$ the formula (\ref{conformalein}) is not
acceptable. We will consider this case in the next subsection.

\subsection{Radiation case}
\label{rhoradiation} Up to now we have not obtained any power-law
solution in the Einstein frame for the case when $w_\mathrm{m}=1/3$,
i.e. with a radiation sector. There, using the conservation
law~(\ref{conserEfin}), we find
\begin{equation}\label{rhosub1}
\rho(t_E)\propto t_E^{-4n_E}\,,
\end{equation}
while Eq.~(\ref{einEOMspe}) with $\Lambda=0$ yields
\begin{equation}
\label{rhosub2} \rho_E(t_E)=\frac{3n_E(3n_E-1)}{\kappa^2t_E^2}\,,
\end{equation}
which implies that
\begin{equation}\label{rhosubresult1}
n_E=\frac{1}{2},\qquad\rho_E(t_E)=\frac{3}{4\kappa^2t_E^2}\,.
\end{equation}

However, since $\phi(t_E)$ couples with $\psi(t_E)$, we cannot solve
these equations directly from the EOM
(\ref{einEOM-1.2})--(\ref{einsca2-1.2}) in the Einstein frame.
Recall now that, having obtained the solutions (\ref{psisoln}) and
(\ref{xisoln}) in the Jordan frame we can, in principle, conformally
transform  both into their corresponding forms in the Einstein
frame. To achieve this goal, we first note that, using
Eq.~(\ref{nEin}), we can find the power-index in the Jordan frame
$n$ from the corresponding one in the Einstein frame $n_E=1/2$,
as\footnote{Note that, for $n_E=1/2$ and $w_\mathrm{m}=1/3$, we cannot
use (\ref{nin2}) to define $n$, because both numerator and
denominator in this formula are equal to zero.}
\begin{eqnarray}\label{njne}
n=\frac{1}{3(1-\alpha)}\,.
\end{eqnarray}
Inserting this into Eq.~(\ref{conformalfact}), we then get
$\phi(t_E)$ as
\begin{eqnarray}\label{phiein2}
\phi(t_E)=\frac{1-3\alpha}{6\alpha-4}\ln\left(\frac{t_E}{\widetilde{t}_2}\right)\,,
\end{eqnarray}
where $\widetilde{t}_2$ is an integration constant. And inserting
this solution into Eq.~(\ref{equsual}), one obtains the solution for
$\psi(t_E)$:
\begin{eqnarray}\label{psiein2}
\psi(t_E)=\widetilde{\psi}_2\left(\frac{t_E}{\widetilde{t}_3}\right)^{\frac{3\alpha}{4-6\alpha}}+\frac{3\alpha-1}
{\alpha\left(3\alpha-2\right)}\ln\left(\frac{t_E}{\widetilde{t}_3}\right)\,,
\end{eqnarray}
with two integration constants, $\widetilde{\psi}_2$ and
$\widetilde{t}_3$. Eqs.~(\ref{einEOM-1.2})--(\ref{einsca2-1.2})
introduce constraints on these integration constants, namely
\begin{eqnarray}
\widetilde{\psi}_2&=&0\,,\label{constr2-1}\\
\frac{\widetilde{t}_2}{\widetilde{t}_3}&=&\left[\frac{1}{f_0}\left(1-\frac{3\alpha}{2}\right)
\right]^{1+\frac{1}{1-3\alpha}}\,.\label{constr2-2}
\end{eqnarray}
Thus, Eqs.~(\ref{rhosubresult1}), (\ref{phiein2})--(\ref{constr2-2})
supplement the solutions in Sect.~\ref{lambdacase}.

\subsection{The case $\Lambda\neq 0$}
\label{nonzerolam}

In the case $\Lambda\neq 0$ solutions in the Jordan frame are
described by  (\ref{psisoln}) and (\ref{xisoln}). To get the
corresponding power-law solutions in the Einstein frame we need to
select the case $m\neq 1-3n$ and to put $\xi_0=-1$. From
(\ref{nonvanishlamb}), we get that the system does not include
matter: $\rho_0=0$. It is easy to see that, for any nonzero $n$,
$C_1\neq 0$ for $m=2$, so we put $\xi_1=0$ and consider the case
$C_2=0$. From (\ref{omegageneral}), we obtain
\begin{equation}
\Omega(t)=\frac{1}{\sqrt{C_1}t}.
\end{equation}
Therefore,
\begin{equation}
t_E=\frac{\sqrt{C_1}}{2}t^2, \qquad H_E=\frac{n+1}{2t_E}, \qquad
n_E=\frac{n+1}{2},
\end{equation}
and
\begin{equation}
\phi=-\ln(\sqrt{C_1}t)=-\frac{1}{2}\ln\left(2\sqrt{C_1}t_E\right)=
\frac{1}{4}\ln\left[\frac{(3n_E-1)n_E}{\Lambda t_E^2}\right]\,,
\end{equation}
where Eq.~(\ref{define}) is used in the last step. Thus, we reobtain the
solution (\ref{phisol1}). Using (\ref{n_alpha}), we reobtain the
condition (\ref{neinsteinvacuum}). Therefore, in the case of nonzero
$\Lambda$ we can use the  power-law solutions of the Jordan frame to
get the corresponding solutions in the Einstein frame, but this way
is here not more effective than a straightforward  search for
power-law solutions of the system
(\ref{einEOM-1.2})--(\ref{einsca2-1.2}).

\subsection{Brief summary of the solutions in the Einstein frame}
\label{eintable}

To render it easier for readers to examine the full
set of solutions, in Tables~\ref{t2f1}, \ref{t2f2} and \ref{t2f3} we
list all those corresponding to the Einstein frame. It should be
noted that all solutions in these three tables correspond to the case
$\Lambda=0$.  As discussed in Sects.~\ref{Einvacuum} and
\ref{nonzerolam}, for the non-vanishing cosmological constant case
in the Einstein frame only vacuum solution can be found, namely
those of Eqs.~(\ref{neinsteinvacuum}), (\ref{phisolution}) and
(\ref{psisolution}).
\begin{table}[tbp]
\centering
\begin{tabular}{l@{}ll@{}ll@{}l}
\hline \multicolumn{2}{c}{solutions}&\multicolumn{2}{c}{constraints}
\\ \hline
&&&&\\
&$\begin{cases}&\displaystyle{\phi(t_E)=\frac{2-3n_E(1+w_\mathrm{m})}{3w-1}\ln\left(\frac{t_E}{\widetilde{t}_1}\right)}\,,\\
\\
&\displaystyle{\psi(t_E)=\frac{
12(2n_E-1)\left[3(1-w_\mathrm{m})+n_E(3w_\mathrm{m}-5)\right]}
{\left[5-3w_\mathrm{m}+3n_E(w_\mathrm{m}-3)\right](3w_\mathrm{m}-1)}\ln\left(\frac{t_E}{\widetilde{t}_2}\right)}\,,\\
\\
&\displaystyle{\rho_E(t_E)=\frac{2n_E(3n_E-1)}{\kappa^2(1-w_\mathrm{m})t_E^2}}\,,\end{cases}$&
&$\begin{cases} \ \mbox{\rm Eq.}\, (\ref{constr1-2})\\ \ \mbox{\rm Eq.}\, (\ref{constr1-3})\\ \ \mbox{\rm Eq.}\, (\ref{alconstr})\end{cases}$&\\
&&&&\\
\hline
\end{tabular}
\caption{Solutions in the Einstein frame for $\displaystyle n_E\neq
1/3$, $\displaystyle{w_\mathrm{m}\neq1, 1/3}$}\label{t2f1}
\end{table}

\begin{table}[tbp]
\centering
\begin{tabular}{l@{}ll@{}ll@{}l}
\hline
\multicolumn{2}{c}{solutions}&\multicolumn{2}{c}{constraints (Eqs.~(\ref{finalver}) and (\ref{constre2}))}\\
\hline
&&&&\\
&$\begin{cases}&\displaystyle{\phi(t_E)=\frac{3n_0-1}{3(1-n_0)}\ln\left(\frac{t_E}{\widetilde{t}_3}\right)}\,, \\
\\
&\displaystyle{\psi(t_E)=\frac{4n_0(2n-1)}{(n_0-1)(3n_0-1)}
\ln\left(\frac{t_E}{\widetilde{t}_4}\right)}\,,\\
\\
&\displaystyle{\rho(t_E)=\widetilde{\rho}_3
t_E^{\frac{4}{3(n_0-1)}}}\,,\end{cases}$&
&$\qquad\begin{cases}\displaystyle{\widetilde{\rho}_3=\frac
{8{n_0}\,{f_0}\, \left( 1-2n_0 \right) }{3 \kappa^2 \left(
1-{n_0}\right)^2
}\left(\frac{\widetilde{t}_3}{\widetilde{t}_4^{\,2}}\right)^{\frac{2(3n_0-1)}{3(n_0-1)}}}\,,\\
\\
\displaystyle{n_0=\frac{-6+3\alpha\pm\sqrt{3\alpha(3\alpha-4)}}{6(2\alpha-3)}}\,.\end{cases}$\\
\hline
\end{tabular}
\caption{Solutions in the Einstein frame for $\displaystyle
n_E=1/3$, $\displaystyle{w_\mathrm{m}=1}$}\label{t2f2}
\end{table}

\begin{table}[tbp]
\centering
\begin{tabular}{l@{}ll@{}ll@{}l}
\hline
\multicolumn{2}{c}{solutions}&\multicolumn{2}{c}{constraints (\mbox{\rm Eq.}\,~(\ref{constr2-2}))}\\
\hline
&&&&\\
&$\begin{cases}&\displaystyle{\phi(t_E)=\frac{1-3\alpha}{6\alpha-4}\ln\left(\frac{t_E}{\widetilde{t}_5}\right)}\,,\\
\\
&\displaystyle{\psi(t_E)=\frac{3\alpha-1}
{\alpha\left(3\alpha-2\right)}\ln\left(\frac{t_E}{\widetilde{t}_6}\right)}\,,\\
\\
&\displaystyle{\rho_E(t_E)=\frac{3}{4\kappa^2t_E^2}}\,,\end{cases}$&
&$\qquad\displaystyle{\frac{\widetilde{t}_2}{\widetilde{t}_3}=\left[\frac{1}{f_0}\left(1-\frac{3\alpha}{2}\right)
\right]^{1+\frac{1}{1-3\alpha}}}\,.$\\[7.2mm]
&&&&\\
\hline
\end{tabular}
\caption{Solutions in the Einstein frame for $\displaystyle
n_E=1/2$, $\displaystyle{w_\mathrm{m}=1/3}$}\label{t2f3}
\end{table}

\section{Conclusions}
\label{conclude} In General Relativity, power-law solutions of the
type $H=n/t$ correspond to models with a perfect fluid whose EoS
parameter $w_{\mathrm{m}}\equiv P_{\mathrm{m}}/\rho_{\mathrm{m}}$ is
related to the power-index by $w_{\mathrm{m}}=-1+2/(3n)$. It is
interesting to try to find similar power-law solutions in modified
gravity theories, in order to check how much they deviate  from
those for GR. In this paper, we consider power-law solutions in a
class of nonlocal gravity models stemming from the widely probed and
very promising function $f(\Box^{-1}R)=f_0e^{\alpha\Box^{-1}R}$
and which include a perfect fluid with constant EoS parameter
$w_{\mathrm{m}}$.

By recasting the original nonlocal action (\ref{action}) into its
local presentation (\ref{action2}), we have obtained power-law
solutions for this model, with and without a cosmological constant
and both in the Jordan and in the Einstein frames. We also show that power-law solutions, obtained in the Jordan frame satisfy the original nonlocal equations. In other words we get power-law solutions for the  original nonlocal model (\ref{action}) as well. In the Jordan
frame we have reached the remarkable conclusion that all power-law
solutions could be found (see Sect.~\ref{jordan}), what is a most
interesting outcome of this paper. In the Einstein frame, we have
correspondingly obtained the power-law solutions either by directly
solving the EOM, or by performing a conformal transformation of the
solutions obtained in the Jordan frame. For this purpose, we have
generalized the correspondence between power-law solutions in the
Jordan and Einstein frames, as obtained in~\cite{ZS}, in order to
appropriately include the matter sector. By using this powerful, non-trivial
tool, we have obtained the solutions when $w_{\mathrm{m}}=1/3$ and
$w_{\mathrm{m}}=1$, in which cases it was very difficult to obtain
the corresponding solutions by directly solving the system
(\ref{einEOM-1.2})--(\ref{einsca2-1.2}). Hence, we have shown
explicitly how the construction of solutions by using conformal
transformation between the two frames proceeds, thus proving
that the method offers a valuable alternative in the search for new
solutions.

In~\cite{EPV2012}, it has been shown that not only models with
exponential $f(\psi)$ can have power-law and de Sitter solutions. It
would be interesting to consider power-law and de Sitter solutions
in the models where $f(\psi)$ consists of a sum of exponentials.
Another direct generalization of the present analysis is to include
several perfect fluid components with different constant values
of $w_{\mathrm{m}}$.

In~\cite{Bamba1104,EPV2011}, de Sitter solutions in nonlocal models
were found. It will be interesting to check the possibilities for the
Universe evolution as obtained from these models, from an
inflationary de Sitter stage to the late power-law Universe and,
furthermore, to check for deviations from the standard general relativity case, and
its distinction from other modified gravity theories.

As is widely known, theories with higher derivatives often
suffer from a ghost problem, namely a wrong sign in the kinetic
term, resulting in a dangerous instability problem. A good aspect in
making use of the conformal transformation technique between the two
frames is to obtain the corresponding ghost-free
conditions~\cite{Nojiri:2010pw,Bamba1104,ZS}.
The biscalar-tensor representation introduces two extra
scalars. As pointed out in~\cite{Tsamis:2010pt,Nojiri:2010pw}, they can lead to a ghost
problem. We should note that the equivalence between the initial
nonlocal theory and local formulation has not been
established yet. The original
nonlocal model has less degrees of freedom and, thus, the
ghost-like behavior of the biscalar-tensor theory may not be a
physical problem, since the associated terms can be cast as boundary
terms of the nonlocal operators~\cite{Koivisto:2010}.
In other words, the would-be ghost mode might not be physically relevant since
it would probably correspond to an inappropriate choice on the
boundary condition. Anyway, it should be kept in mind that an
appropriate choice of boundary condition is also necessary in the
biscalar-tensor presentation. We plan to consider this important question in
more detail in further work on the original nonlocal model.

It will be most interesting, too,
to test the solutions obtained in this paper to find the constraints
on the parameters, hence to check for the possibility to obtain a
realistic model which can be responsible for the current
observed acceleration of the Universe expansion. This work is
now in process. Also, an analysis of the stability of the solutions
here encountered will be carried out~\cite{LSZ:2013}.

\medskip

\noindent {\bf Acknowledgements.} The authors thank
Sergei~D.~Odintsov and Misao~Sasaki for very useful discussions.
E.E. was supported in part by MINECO (Spain), grant PR2011-0128 and
project FIS2010-15640, by the CPAN Consolider Ingenio Project, and
by AGAUR (Generalitat de Ca\-ta\-lu\-nya), contract 2009SGR-994. His
research was partly carried out while on leave at the Department of
Physics and Astronomy, Dartmouth College, NH, USA. The research of
E.P. and S.V. was supported in part by the RFBR grant 11-01-00894,
E.P. by the RFBR grant 12-02-31109, and S.V. by the Russian Ministry
of Education and Science under grant NSh-3920.2012.2, and by
contract CPAN10-PD12 (ICE, Barcelona, Spain). Y.Z. was supported by
the Grant-in-Aid for the Global COE Program ``The Next Generation of
Physics, Spun from Universality and Emergence'' from the Ministry of
Education, Culture, Sports, Science and Technology (MEXT) of Japan,
by JSPS Grant-in-Aid for Scientific Research (A) No. 21244033.



\begin{thebibliography}{99}


\bibitem{Perlmutter:1998np}
 S.~Perlmutter {\it et al.}  [SNCP Collaboration],
 \textit{Measurements of Omega and Lambda from 42 High-Redshift
 Supernovae},
 \textit{Astrophys.\ J.}  {\bf 517} (1999) 565,
 [arXiv:astro-ph/981213].
%
\bibitem{Riess:1998cb}
  A.G.~Riess {\it et al.}  [Supernova Search Team Collaboration],
  \textit{Observational Evidence from Supernovae for an Accelerating Universe and a
  Cosmological Constant},
  \textit{Astron.\ J.}  {\bf 116} (1998)  1009,
  [arXiv:astro-ph/9805201]; \\
A.G.~Riess {\it et al.}  [Supernova Search Team Collaboration],
\textit{Type Ia Supernova Discoveries at z>1 From the Hubble Space
Telescope: Evidence for Past Deceleration and Constraints on Dark
Energy Evolution}, \textit{Astrophys. J.}
\textbf{607} (2004) 665, [arXiv:astro-ph/0402512]; \\
P. Astier {\it et al.}, \textit{The Supernova Legacy Survey:
Measurement of $\Omega_M$, $\Omega_\Lambda$ and $w$ from the First
Year Data Set}, \textit{Astron. Astrophys.} \textbf{447} (2006) 31,
[arXiv:astro-ph/0510447].

\bibitem{WMAP}
 D.N.~Spergel {\it et al.}  [WMAP Collaboration],
 \textit{First Year Wilkinson Microwave Anisotropy Probe (WMAP) Observations:
 Determination of Cosmological Parameters},
 \textit{Astrophys.\ J.\ Suppl.}  {\bf 148} (2003)  175,
 [arXiv:astro-ph/0302209];  \\
 D.N.~Spergel {\it et al.}  [WMAP Collaboration],
 \textit{Wilkinson Microwave Anisotropy Probe (WMAP) three year results:
 Implications for cosmology},
 \textit{Astrophys.\ J.\ Suppl.} {\bf 170} (2007)
 377, [arXiv:astro-ph/0603449]; \\
 E.~Komatsu {\it et al.}  [WMAP Collaboration],
 \textit{Five-Year Wilkinson Microwave Anisotropy Probe (WMAP)
 Observations:Cosmological Interpretation},
 \textit{Astrophys.\ J.\ Suppl.}  \textbf{180} (2009) {330}
 [arXiv:0803.0547];\\
E.~Komatsu {\it et al.}  [WMAP Collaboration],
  \textit{Seven-Year Wilkinson Microwave Anisotropy Probe (WMAP) Observations:
  Cosmological Interpretation},
  \textit{Astrophys.\ J.\ Suppl.}  {\bf 192} (2011) 18,
  [arXiv:1001.4538].
\bibitem{Tegmark}
  M.~Tegmark {\it et al.}  [SDSS Collaboration],
  \textit{Cosmological parameters from SDSS and WMAP},
  \textit{Phys.\ Rev.}  D {\bf 69} (2004) 103501,
  [arXiv:astro-ph/0310723];  \\
  M. Tegmark \textit{et al.} [SDSS collaboration],
  \textit{The 3D power spectrum of galaxies from the SDS}, \textit{Astroph.~J.}
\textbf{606} (2004) 702--740, [arXiv:astro-ph/0310725].
\bibitem{Seljak:2004xh}
  U.~Seljak {\it et al.}  [SDSS Collaboration],
  \textit{Cosmological parameter analysis including SDSS Ly-alpha forest and  galaxy
  bias: Constraints on the primordial spectrum of fluctuations,
  neutrino mass, and dark energy},
  \textit{Phys.\ Rev.}  D {\bf 71} (2005) 103515,
  [arXiv:astro-ph/0407372].
\bibitem{Eisenstein:2005su}
  D.J.~Eisenstein {\it et al.}  [SDSS Collaboration],
  \textit{Detection of the Baryon Acoustic Peak in the Large-Scale Correlation
  Function of SDSS Luminous Red Galaxies},
  \textit{Astrophys.\ J.}  {\bf 633} (2005) 560,
  [arXiv:astro-ph/0501171].
\bibitem{Jain:2003tba}
  B.~Jain and A.~Taylor,
  \textit{Cross-correlation Tomography: Measuring Dark Energy Evolution with Weak
  Lensing},
  \textit{Phys.\ Rev.\ Lett.}  {\bf 91} (2003) 141302,
  [arXiv:astro-ph/0306046].
\bibitem{Kilbinger:2008gk}
  M.~Kilbinger {\it et al.},
\textit{Dark energy constraints and correlations with systematics
from CFHTLS weak lensing, SNLS supernovae Ia and WMAP5},
\textit{Astron. Astrophys.} \textbf{497} (2009) 677--688,
[arXiv:0810.5129].

\bibitem{QuintomModels}
E. Elizalde, S. Nojiri, S.D. Odintsov, \textit{Late-time cosmology
in (phantom) scalar-tensor theory: dark energy and the cosmic
speed-up},
\textit{Phys.\ Rev.\ D} \textbf{70} (2004) 043539, [arXiv:hep-th/0405034]; \\
Hrv. \u{S}tefan\u{c}i\'{c},
\textit{Dark energy transition between
quintessence and phantom regimes --- an equation of state analysis},
\textit{Phys.\ Rev.\ D} \textbf{71} (2005) 124036, [arXiv:astro-ph/0504518];\\
W. Zhao and Y. Zhang, \textit{The Quintom Models with State Equation
Crossing $-1$},
\textit{Phys.\ Rev.\ D} \textbf{73} (2006) 123509, [arXiv:astro-ph/0604460];\\
 X. Zhang,
\textit{Dynamical vacuum energy, holographic quintom, and the
reconstruction of scalar-field dark energy}, \textit{Phys.\ Rev.\ D}
\textbf{74} (2006) 103505,  [arXiv:astro-ph/0609699].


\bibitem{Guo2004} Zong-Kuan Guo, Yun-Song Piao, Xinmin Zhang, Yuan-Zhong
Zhang, \textit{Cosmological Evolution of a Quintom Model of Dark
Energy}, \textit{Phys.\ Lett.\ B} \textbf{608} (2005) 177,
[arXiv:astro-ph/0410654].

\bibitem{AKV2} I.Ya. Aref'eva, A.S. Koshelev, and S.Yu. Vernov,
\textit{Crossing the $w=-1$ barrier in the D3-brane dark energy
model}, \textit{Phys.\ Rev.\ D} \textbf{72} (2005)
064017, [arXiv:astro-ph/0507067];\\
S.Yu. Vernov, \textit{Construction of Exact Solutions in Two-Fields
Models and the Crossing of the Cosmological Constant Barrier},
\textit{Theor.\ Math.\ Phys.} \textbf{155} (2008) 544--556, [arXiv:astro-ph/0612487];\\
I.Ya. Aref'eva, N.V. Bulatov, and S.Yu. Vernov, \textit{Stable exact
solutions in cosmological models with two scalar fields},
\textit{Theor.\ Math.\ Phys.} \textbf{163} (2010) 788,
[arXiv:0911.5105].

\bibitem{Andrianov}
A.A.~Andrianov, F.~Cannata, A.Yu.~Kamenshchik, and D.~Regoli,
\textit{Reconstruction of scalar potentials in two-field
cosmological models}, \textit{J.\ Cosmol.\ Astropart.\ Phys.} {\bf
0802} (2008) 015, [arXiv:0711.4300].

\bibitem{Lazkoz}
R. Lazkoz, G.~Le\'on, \textit{Quintom cosmologies admitting either
tracking or phantom attractors}, \textit{Phys.\ Lett.\ B}
\textbf{638} (2006)
303, [arXiv:astro-ph/0602590];\\
R.~Lazkoz, G.~Le\'on, and I. Quiros, \textit{Quintom cosmologies
with arbitrary potentials}, \textit{Phys.\ Lett.\ B} \textbf{649}
(2007) 103, [arXiv:astro-ph/0701353].


\bibitem{Setare}
J. Sadeghi, M.R. Setare, A. Banijamali, \textit{String Inspired
Quintom Model with Non-minimally Coupled Modified Gravity},
\textit{Phys.\ Lett.\ B} \textbf{678} (2009) 164--167,
[arXiv:0903.4073];\\
M.R. Setare, E.N. Saridakis, \textit{ Quintom dark energy models
with nearly flat potentials},
\textit{Phys.\ Rev.\ D} \textbf{79} (2009) 043005, [arXiv:0810.4775];\\
M.R. Setare, E.N. Saridakis, \textit{Quintom Cosmology with General
Potentials}, \textit{Int.\ J.\ Mod.\ Phys.\ D} \textbf{18} (2009)
549, [arXiv:0807.3807].

\bibitem{Quinmodrev1}
Y.-F. Cai, E.N. Saridakis, M.R. Setare, and
  J.-Q. Xia, \textit{Quintom Cosmology: theoretical implications and
observations},
\textit{Phys.\ Rep.} \textbf{493} (2010) 1, [arXiv:0909.2776];\\
H. Zhang, \textit{Crossing the phantom divide}, arXiv:0909.3013.


\bibitem{NO-rev}
 S.~Nojiri and S.D.~Odintsov,
 \textit{Introduction to modified gravity and gravitational alternative for dark
 energy},
 \textit{Int.\ J.\ Geom.\ Meth.\ Mod.\ Phys.}  {\bf 4} (2007) 115, arXiv:hep-th/0601213

\bibitem{Review-Nojiri-Odintsov}
  S.~Nojiri and S.D.~Odintsov,
  \textit{Unified cosmic history in modified gravity: from F(R) theory to Lorentz
  non-invariant models},
  \textit{Phys. Rept.} \textbf{505} (2011) 59, [arXiv:1011.0544].

\bibitem{Felice_Tsujikawa}
A. de Felice, Sh. Tsujikawa,
\textit{Living Rev. Rel.} \textbf{13} (2010) 3, arXiv:1002.4928


\bibitem{Fujii_Maeda} Y. Fujii, K. Maeda,  \textit{The Scalar-Tensor Theory of Gravitation}, Cambridge University Press,
Cambridge, 2004


\bibitem{Book-Capozziello-Faraoni}
S.~Capozziello and V.~Faraoni, \textit{Beyond Einstein Gravity: A
Survey of Gravitational Theories for Cosmology and Astrophysics},
\textit{Fund.\ Theor.\ Phys.} \textbf{170}, Springer, New York,
2011.

\bibitem{CL}
S. Capozziello and M. De Laurentis, \textit{Extended Theories of
Gravity}, \textit{Phys.\ Rep.} {\bf 509} (2011) 167,
[arXiv:1108.6266].



\bibitem{BCNO} K. Bamba, S. Capozziello, S. Nojiri, and S.D. Odintsov,
\textit{Dark energy cosmology: the equivalent description via
different theoretical models and cosmography tests},
\textit{Astrophys.\ Space\ Sci.} \textbf{342} (2012) 155,
[arXiv:1205.3421].



\bibitem{Stelle}
K.S.~Stelle,
  \textit{Renormalization of Higher Derivative Quantum Gravity},
  \textit{Phys. Rev.} D \textbf{16} (1977) 953.

\bibitem{BOSh} I. L. Buchbinder,
S.D. Odintsov, and I.L. Shapiro, \textit{Effective Action in Quantum
Gravity}, IOP, Bristol, 1992.

\bibitem{Deser:2007jk}
  S.~Deser and R.~P.~Woodard,
  \textit{Nonlocal Cosmology},
  \textit{Phys.\ Rev.\ Lett.}\  {\bf 99} (2007) 111301,
  [arXiv:0706.2151].

\bibitem{Non-local-gravity-Refs}
 N.~Arkani-Hamed, S.~Dimopoulos, G.~Dvali and G.~Gabadadze,
  \textit{Non-local modification of gravity and the cosmological constant
  problem},
  [arXiv:hep-th/0209227];\\
 D. Espriu, T. Multamaki, and E.C. Vagenas,
\textit{Cosmological significance of one-loop effective gravity},
 \textit{Phys.\ Lett.} B \textbf{628} (2005) 197, [arXiv:gr-qc/0503033];\\
 T. Biswas, A. Mazumdar, and W. Siegel,
\textit{Bouncing Universes in String-inspired Gravity},
\textit{J. Cosmol. Astropart. Phys.} \textbf{0603} (2006) 009, [arXiv:hep-th/0508194]; \\
  G.~Calcagni, M.~Montobbio, and G.~Nardelli,
\textit{Localization of nonlocal theories},
  \textit{Phys.\ Lett.}  B {\bf 662} (2008)  285,
  [arXiv:0712.2237]; \\
%
  S.~Capozziello, E.~Elizalde, S.~Nojiri, and S.D.~Odintsov,
  \textit{Accelerating cosmologies from non-local higher-derivative gravity},
  \textit{Phys. Lett.}  B {\bf 671}  (2009)
 193,  [arXiv:0809.1535]; \\
S.~Nesseris and A.~Mazumdar,
  \textit{Newton's constant in $f(R,R_{\mu\nu}R^{\mu\nu},\Box R)$ theories of gravity
  and constraints from BBN},
  \textit{Phys.\ Rev.}  D  \textbf{79} (2009) 104006
  [arXiv:0902.1185]; \\
F.W. Hehl and  B. Mashhoon,
 \textit{A formal framework for a nonlocal
generalization of Einstein's theory of gravitation}, \textit{Phys.
Rev.} D \textbf{79} (2009) 064028, [arXiv:0902.0560]; \\
  G.~Calcagni and G.~Nardelli,
  \textit{Cosmological rolling solutions of nonlocal theories},
  \textit{Int.\ J.\ Mod.\ Phys.}  D {\bf 19} (2010) 329,
  [arXiv:0904.4245]; \\
  G.~Cognola, E.~Elizalde, S.~Nojiri, S.D.~Odintsov, and S.~Zerbini,
  \textit{One-loop effective action for non-local modified Gauss-Bonnet gravity
  in deSitter space},
 \textit{Eur.\ Phys.\ J.}  C {\bf 64} (2009) 483,
  [arXiv:0905.0543]; \\
  G.~Calcagni and G.~Nardelli,
  \textit{Nonlocal gravity and the diffusion equation},
  \textit{Phys. Rev.} D {\bf  82} (2010) 123518,
  [arXiv:1004.5144]; \\
  T.~Biswas, T.~Koivisto, and A.~Mazumdar,
  \textit{Towards a resolution of the cosmological singularity in non-local higher
  derivative theories of gravity},
  \textit{J.\ Cosmol.\ Astropart.\ Phys.} {\bf 1011} (2010) 008, [arXiv:1005.0590]; \\
L. Modesto, \textit{Super-renormalizable Quantum Gravity},
\textit{Phys.\ Rev.} D \textbf{86} (2012) 044005, [arXiv:1107.2403]; \\
A.O.~Barvinsky, \textit{Serendipitous discoveries in nonlocal
gravity theory},
  \textit{Phys. Rev.} D {\bf 85} (2012) 104018,
  [arXiv:1112.4340];\\
A.S. Koshelev and S.Yu. Vernov, \textit{On bouncing solutions in
non-local gravity},
\textit{Phys. Part. Nucl.} \textbf{43} (2012) 666, [arXiv:1202.1289];\\
T. Biswas, A.S. Koshelev, A. Mazumdar, and S.Yu. Vernov,
\textit{Stable bounce and inflation in non-local higher derivative
cosmology},
\textit{J. Cosmol. Astropart. Phys.} \textbf{1208} (2012) 024, [arXiv:1206.6374];\\
A.S. Koshelev, \textit{Stable analytic bounce in non-local
Einstein-Gauss-Bonnet cosmology}, arXiv:1302.2140.

\bibitem{Odintsov0708}
  S.~Nojiri and S.D.~Odintsov,
  \textit{Modified non-local-F(R) gravity as the key for the inflation and dark
  energy,}
  \textit{Phys.\ Lett.}  B {\bf 659} (2008)
  821,  [arXiv:0708.0924].


\bibitem{Jhingan:2008ym}
  S.~Jhingan, S.~Nojiri, S.D.~Odintsov, M.~Sami, I.~Thongkool, and S.~Zerbini,
  \textit{Phantom and non-phantom dark energy: The cosmological relevance of
  non-locally corrected gravity},
  \textit{Phys.\ Lett.}  B {\bf 663} (2008) 424, [arXiv:0803.2613].

\bibitem{Koivisto:2008}
T.S.~Koivisto,
  \textit{Dynamics of Nonlocal Cosmology},
  \textit{Phys.\ Rev.}  D {\bf 77} (2008) 123513,
  [arXiv:0803.3399].

\bibitem{Koivisto2}
T.S.~Koivisto,
  \textit{Newtonian limit of nonlocal cosmology},
\textit{Phys.\ Rev.}  D {\bf 78} (2008) 123505, [arXiv:0807.3778].

\bibitem{Woodard}
  C.~Deffayet and R.P.~Woodard,
  \textit{Reconstructing the Distortion Function for Nonlocal
  Cosmology},
  \textit{J.\ Cosmol.\ Astropart.\ Phys.} {\bf 0908} (2009) 023,
  [arXiv:0904.0961].


\bibitem{Non-local-FR}
  N.A.~Koshelev,
  \textit{Comments on scalar-tensor representation of nonlocally
  corrected gravity},
  \textit{Grav.\ Cosmol.} {\bf 15}  (2009) 220,
  [arXiv:0809.4927]; \\
  K.A.~Bronnikov and E.~Elizalde,
  \textit{Spherical systems in models of nonlocally corrected
  gravity},
  \textit{Phys.\ Rev.}  D {\bf 81} (2010) 044032,
  [arXiv:0910.3929]; \\
A.J. L\'opez-Revelles and E. Elizalde, \textit{Universal procedure
to cure future singularities of dark energy models},
\textit{Gen. Rel. Grav.} {\bf 44} (2012) 751, [arXiv:1104.1123];\\
J.~Kluson,
 \textit{Non-Local Gravity from Hamiltonian Point of View},
 \textit{JHEP} {\bf 1109} (2011) 001, [arXiv:1105.6056]
\bibitem{Tsamis:2010pt}
  N.C.~Tsamis and R.P.~Woodard,
  \textit{Primordial Density Perturbations and Reheating from Gravity},
  Phys.\ Rev.\ D {\bf 82} (2010) 063502, [arXiv:1006.4834]
\bibitem{ParkDonelson}
S. Park and S. Dodelson, \textit{Structure formation in a nonlocally
modified gravity model}, \textit{Phys.\ Rev.} D \textbf{87} (2013)
024003, [arXiv:1209.0836]


\bibitem{Nojiri:2010pw}
  S.~Nojiri, S.D.~Odintsov, M.~Sasaki and Y.l.~Zhang,
  \textit{Screening of cosmological constant in non-local gravity},
  \textit{Phys.\ Lett.}  B {\bf 696} (2011) 278,
  [arXiv:1010.5375].


\bibitem{Bamba1104}
K. Bamba, S. Nojiri, S.D. Odintsov, and M. Sasaki, \textit{Screening
of cosmological constant for De Sitter Universe in non-local
gravity, phantom-divide crossing and finite-time future
singularities}, \textit{General Relativity and Gravitation}
\textbf{44} (2012) 1321, [arXiv:1104.2692].


\bibitem{ZS} Y.l.~Zhang and M.~Sasaki,
\textit{Screening of cosmological constant in non-local cosmology},
\emph{Int.\ J.\ Mod.\ Phys.\ D} \textbf{21} (2012) 1250006,
[arXiv:1108.2112].

\bibitem{EPV2011}
 E. Elizalde, E.O. Pozdeeva, and S.Yu. Vernov,
\textit{De Sitter Universe in Non-local Gravity}, \textit{Phys.
Rev.} D \textbf{85} (2012) 044002, [arXiv:1110.5806];
\bibitem{Vernov1202} S.Yu. Vernov,
\textit{Nonlocal Gravitational Models and Exact Solutions},
\textit{Phys.\ Part.\ Nucl.} \textbf{43} (2012) 694,
[arXiv:1202.1172];
\bibitem{EPV2012}
E. Elizalde, E.O. Pozdeeva, and S.Yu. Vernov, \textit{Reconstruction
Procedure in Nonlocal Models}, \textit{Class.\ Quantum\ Grav.}
\textbf{30} (2013) 035002, [arXiv:1209.5957].


\bibitem{Non-local_scalar}
I.Ya. Aref'eva, \textit{Nonlocal String Tachyon as a Model for
Cosmological Dark Energy}, \textit{AIP Conf. Proc.} \textbf{826},
\textit{p-Adic Mathematical Physics}, eds. A.Yu. Khrennikov,
Z. Rakich, I.V. Volovich, AIP, Melville, NY, 2006, pp.~301--311, [arXiv:astro-ph/0410443]; \\
G. Calcagni, \textit{Cosmological tachyon from cubic string field
theory},
 \textit{JHEP} \textbf{0605} (2006) 012, [arXiv:hep-th/0512259]; \\
A.S. Koshelev,
 \textit{Non-local SFT Tachyon and Cosmology},
 \textit{JHEP}
\textbf{0704} (2007) 029, [arXiv:hep-th/0701103]; \\
I.Ya. Aref'eva, L.V. Joukovskaya, and S.Yu. Vernov,
 \textit{Bouncing and accelerating solutions in nonlocal stringy models}
 \textit{JHEP} \textbf{0707} (2007) 087, [arXiv:hep-th/0701184]; \\
 I.Ya. Aref'eva and  I.V.  Volovich,
 \textit{Quantization of the Riemann
 Zeta-Function and Cosmology},
 \textit{Int.\ J.\ of\ Geom.\ Meth.\
Mod.\ Phys.}  \textbf{4} (2007) 881, [arXiv:hep-th/0701284]; \\
 J.E. Lidsey,
 \textit{Stretching the Inflaton Potential with Kinetic Energy},
 \textit{Phys.\ Rev.}  D {\bf 76} (2007) 043511, [arXiv:hep-th/0703007]; \\
 G. Calcagni, M. Montobbio, and  G. Nardelli,
 \textit{A route to nonlocal cosmology},
\textit{Phys.\ Rev.} D \textbf{76} (2007) 126001,
[arXiv:0705.3043]; \\
N. Barnaby, T. Biswas, and  J.M. Cline, \textit{p-adic Inflation},
\textit{JHEP} \textbf{0704} (2007) 056,
 [arXiv:hep-th/0612230]; \\
 N. Barnaby and  J.M. Cline,
 \textit{Large Nongaussianity from Nonlocal Inflation},
 \textit{J. Cosmol. Astropart. Phys.} \textbf{0707} (2007) 017,
[arXiv:0704.3426]; \\
I.Ya. Aref'eva, L.V. Joukovskaya, and S.Yu. Vernov,
 \textit{Dynamics in nonlocal linear models in the Friedmann--Robertson--Walker metric},
 \textit{J. Phys. A: Math. Theor.} \textbf{41} (2008) 304003,
[arXiv:0711.1364]; \\
D.J. Mulryne and N.J. Nunes, \textit{Diffusing non-local inflation:
Solving the field equations as an initial value problem},
\textit{Phys.\ Rev.}  D \textbf{78} (2008) 063519, [arXiv:0805.0449]; \\
G. Calcagni, M. Montobbio, and  G. Nardelli,
\textit{Phys.\ Lett.} B \textbf{662} (2008) 285, [arXiv:0712.2237]; \\
G. Calcagni and  G. Nardelli, \textit{Nonlocal instantons and
solitons in string models},
\textit{Phys.\ Lett.} B \textbf{669} (2008) 102, [arXiv:0802.4395]; \\
L.V. Joukovskaya, \textit{Dynamics in nonlocal cosmological models
derived from string field theory}
  \textit{Phys. Rev.} D \textbf{76} (2007) 105007, [arXiv:0707.1545]; \\
L.V. Joukovskaya,
 \textit{Dynamics with Infinitely Many Time Derivatives in Friedmann--Robertson--Walker Background
  and Rolling Tachyon},
  \textit{JHEP} \textbf{0902} (2009) 045, [arXiv:0807.2065];\\
N. Barnaby and  N. Kamran, \textit{Dynamics with Infinitely Many
Derivatives: The Initial Value Problem},
\textit{JHEP} {\bf 0802} (2008) 008, [arXiv:0709.3968]; \\
G. Calcagni and G. Nardelli, \textit{Cosmological rolling solutions
of nonlocal theories}, \textit{Int.\ J.\ Mod.\ Phys.} D \textbf{19}
(2010) 329, [arXiv:0904.4245]; \\
S.Yu. Vernov, \textit{Localization of nonlocal cosmological models
with quadratic potentials in the case of double roots},
\textit{Class.\ Quant.\ Grav.} \textbf{27} (2010) 035006, [arXiv:0907.0468]; \\
S.Yu.~Vernov,
 \textit{Localization of the SFT inspired Nonlocal Linear Models and Exact Solutions},
\textit{Phys. Part. Nucl. Lett.} \textbf{8} (2011) 310, [arXiv:1005.0372]; \\
A.S. Koshelev and S.Yu. Vernov, \textit{Cosmological perturbations
in SFT inspired non-local scalar field models},
\textit{Eur.\ Phys.\ J.} C \textbf{72} (2012) 2198, [arXiv:0903.5176]; \\
A.S. Koshelev and S.Yu. Vernov, \textit{Analysis of scalar
perturbations in cosmological models with a non-local scalar field},
\textit{Class. Quant. Grav.} \textbf{28} (2011) 085019, [arXiv:1009.0746]; \\
  N.~Barnaby,
  \textit{A New Formulation of the Initial Value Problem for Nonlocal Theories},
  \textit{Nucl.\ Phys.}\  B {\bf 845} (2011) 1, [arXiv:1005.2945].


\bibitem{BD:1961}
C. Brans and R.H. Dicke, \textit{Mach's principle and a relativistic
theory of gravitation},
\textit{Phys.\ Rev.} \textbf{124} (1961) 925;\\
R.H. Dicke, \textit{Mach's principle and invariance under
transformation of units},
\textit{Phys. Rev.} \textbf{125} (1962) 2163;\\
M. Fierz, \textit{On the physical interpretation of P.Jordan's
extended theory of gravitation}, \textit{Helv. Phys. Acta }
\textbf{29} (1956) 128.

\bibitem{Faraoni_OLD} V. Faraoni, E. Gunzig, and P. Nardone,
\textit{Conformal transformations in classical gravitational
theories and in cosmology}, \textit{Fundam.\ Cosm.\ Phys.}
\textbf{20} (1999) 121, [arXiv:gr-qc/9811047].

\bibitem{Faraoni-Nadeau} V. Faraoni and Sh. Nadeau,
\textit{(Pseudo)issue of the conformal frame revisited},
\textit{Phys. Rev.} D \textbf{75} (2007) 023501.

\bibitem{CNOT} S. Capozziello, S. Nojiri, S.D. Odintsov and A. Troisi,
\textit{Cosmological viability of f(R)-gravity as an ideal fluid and
its compatibility with a matter dominated phase},
\textit{Phys.\ Lett.} B \textbf{639} (2006) 135, [arXiv:astro-ph/0604431];\\
C. Corda, \textit{Gravitational wave astronomy: the definitive test
for the ``Einstein frame versus Jordan frame" controversy},
\textit{Astropart.\ Phys.} \textbf{34} (2011) 412, [arXiv:1010.2086];\\
A. Bhadra, K. Sarkar, D.P. Datta and K.K. Nandi, \textit{Brans-Dicke
theory: Jordan vs Einstein Frame}, \textit{Mod.\ Phys.\ Lett.} A
\textbf{22} (2007) 367, [arXiv:gr-qc/0605109].
\bibitem{Capozziello}
S. Capozziello, P. Martin-Moruno and C. Rubano, \textit{Physical
non-equivalence of the Jordan and Einstein frames}, \textit{Phys.
Lett.} B \textbf{689} (2010) 117, [arXiv:1003.5394].

\bibitem{NS:2010a}
N. Deruelle and M. Sasaki, \textit{Conformal equivalence in
classical gravity: the example of "veiled" General Relativity}, in
\textit{Proceedings: Cosmology, Quantum Vacuum and Zeta Functions,
8-10 March, 2010}, [arXiv:1007.3563];\\
N. Deruelle and M. Sasaki, \textit{Inflation with a Weyl term, or
ghosts at work},
\textit{J.\ Cosmol.\ Astropart.\ Phys.} \textbf{1103} (2011) 040, [arXiv:1012.5386];\\
I. Quiros, R. Garcia-Salcedo, J.E.M. Aguilar and T. Matos,
\textit{The conformal transformation's controversy: what are we
missing?}, \textit{Gen.\ Rel.\ Grav.} \textbf{45} (2013) 489,
[arXiv:1108.5857].


\bibitem{Will:2005va}
  C.M.~Will,
  \textit{The Confrontation between General Relativity and Experiment},
  \textit{Living Rev.\ Rel.}  {\bf 9} (2006) 3,
  [arXiv:gr-qc/0510072].

\bibitem{Koivisto:2010}
T.S. Koivisto, \textit{Cosmology of modified (but second order)
gravity},  \textit{AIP Conf.\ Proc.} \textbf{1206} (2010) 79,
[arXiv:0910.4097].


\bibitem{LSZ:2013}
L. Alabidi, M. Sasaki and Y.l. Zhang, in preparation.

\end{thebibliography}
\end{document}